\documentclass[preprint,5p,sort&compress,times,twocolumn]{elsarticle}
\usepackage{amsmath}
\usepackage{graphicx,subfigure,color}

\journal{Physics Letters B}

\begin{document}

\begin{frontmatter}

\title{Stationary solutions for fast flavor oscillations of a homogeneous dense neutrino gas}

\author[address1,address2]{Zewei Xiong}
\ead{z.xiong@gsi.de}
\author[address1]{Yong-Zhong Qian}
\address[address1]{School of Physics and Astronomy, University of Minnesota, Minneapolis, Minnesota 55455, USA}
\address[address2]{GSI Helmholtzzentrum f\"{u}r Schwerioneneforschung, 64291 Darmstadt, Germany}

\begin{abstract}
We present a method to find the stationary solutions for fast flavor oscillations of a homogeneous
dense neutrino gas. These solutions correspond to collective precession of all neutrino polarization vectors 
around a fixed axis in the flavor space on average, and are conveniently studied in the co-rotating frame. 
We show that these solutions can account for the numerical results of explicit evolution calculations, and 
that even with the simplest assumption of adiabatic evolution, they can provide the average survival 
probabilities to good approximation. We also discuss improvement of these solutions and their use as
estimates of the effects of fast oscillations in astrophysical environments.
\end{abstract}

\begin{keyword}

Neutrino oscillations \sep Dense neutrino gas \sep Fast flavor oscillations

\end{keyword}

\end{frontmatter}

\section{Introduction}
\label{sec:introduction}
Experiments with solar, atmospheric, reactor, and accelerator neutrinos have established
that neutrinos produced in a specific flavor state oscillate among all three flavor states \cite{olive2014review}.
Neutrino oscillations studied by the above experiments depend on the vacuum
neutrino mixing properties and forward neutrino-electron scattering in the relevant matter.
Because flavor evolution of individual neutrinos can be treated separately in these cases,
the theory is well understood and calculations are straightforward. In astrophysical 
environments such as supernovae and neutron star mergers, however, neutrino emission is 
so intense that forward scattering among neutrinos becomes important, which causes flavor 
evolution to be coupled for neutrinos emitted with different energies and directions.
Consequently, collective flavor oscillations (see e.g., \cite{duan2010collective} for a review of the early studies) 
may occur for the dense neutrino gas in these environments. The treatment and understanding 
of such oscillations are still evolving and under intensive study
(see e.g., \cite{raffelt2013axial,chakraborty2016collective,izaguirre2017fast,yi2019dispersion,martin2019nonlinear,martin2020spectral}
for some recent developments).

Many insights can be obtained by studying oscillations of astrophysical neutrinos in terms
of mixing between $\nu_e$ and $\nu_x$, which represents an appropriate linear combination of
$\nu_\mu$ and $\nu_\tau$. In this case, the flavor field can be conveniently described by
the polarization vector $\mathsf{P}(\omega,\mathbf{v})$ for a neutrino (antineutrino) emitted
with energy $E$ and velocity $\bf{v}$. Here $\omega=\pm\delta m^2/(2E)$ is the vacuum
oscillation frequency, $\delta m^2>0$ is the mass-squared difference between the two 
vacuum mass eigenstates, and $\omega>0$ ($\omega<0$) denotes neutrinos (antineutrinos).
For clarity, the sans serif font and numeral indices are used for quantities in the flavor space,
while the bold face and $(x,y,z)$ indices are used for vectors in the Euclidean space. 
The probability of being the electron flavor is $(P_3+1)/2$, where 
$P_3=\mathsf{P}\cdot \hat{\mathsf{e}}_3$ and $\hat{\mathsf{e}}_3$ is the unit vector 
in the third direction of the flavor space. For coherent flavor evolution 
(i.e., in the absence of collisions) that starts with all neutrinos 
in pure flavor states, the $\mathsf{P}(\omega,\mathbf{v})$ for the initial $\nu_e$ and $\nu_x$ 
($\bar\nu_e$ and $\bar\nu_x$) will differ only by an overall sign. Hereafter, 
$\mathsf{P}(\omega,\mathbf{v})$ specifically refers to the polarization vector of the 
initial $\nu_e$ or $\bar\nu_e$.

The general equation governing the spatial and temporal evolution of $\mathsf{P}(\omega,\mathbf{v})$ is
\begin{equation}
	(\partial_t+\mathbf{v}\cdot\nabla)\mathsf{P}(\omega,\mathbf{v}) = \mathsf{H}(\omega,\mathbf{v}) \times \mathsf{P}(\omega,\mathbf{v}),
	\label{eq:pvw_vdrho_P}
\end{equation}
where $\mathsf{H}(\omega,\mathbf{v}) = \mathsf{H}_\text{vac}(\omega) + \mathsf{H}_\text{mat} + \mathsf{H}_{\nu\nu}(\mathbf{v})$
is the total Hamiltonian. The term $\mathsf{H}_\text{vac}(\omega)=\omega \mathsf{B}$ accounts for the vacuum mixing,
where $\mathsf{B}=(\sin 2\theta_V,\,0,\,-\cos 2\theta_V)$ for the normal mass hierarchy and 
$\mathsf{B}=(-\sin 2\theta_V,\,0,\,\cos 2\theta_V)$ for the inverted mass hierarchy with $\theta_V$ being the vacuum mixing angle.
The contribution $\mathsf{H}_\text{mat} = \lambda \hat{\mathsf{e}}_3$ originates from forward neutrino-electron scattering, where
$\lambda=\sqrt{2} G_F n_e$, $G_F$ is the Fermi coupling constant, and $n_e$ is the net electron number density.
The contribution $\mathsf{H}_{\nu\nu}(\mathbf{v})$ from forward scattering among neutrinos is our main concern and is discussed below.

We write $\mathsf{H}_{\nu\nu}(\mathbf{v})=v^\rho(\mathbf{v}) \mathsf{J}_\rho$, where $v^\rho(\mathbf{v})=(1, \mathbf{v})$ is the
four-vector corresponding to $\bf{v}$ with $\rho$ running over $(t,x,y,z)$,
\begin{equation}
	\mathsf{J}^\rho = \mu \int d\mathbf{v}'\, v^\rho(\mathbf{v}') \int_{-\infty}^{+\infty} d\omega\, G(\omega,\mathbf{v}') \mathsf{P}(\omega,\mathbf{v}')
	\label{eq:pvw_Jmu}
\end{equation}
is the neutrino polarization current, $\mu=\sqrt{2} G_F n_{\nu_e}$, 
$n_{\nu_e}=\int d\mathbf{v}\, \int_0^{\infty} d\omega\, F_{\nu_e}(\omega,\mathbf{v})$ is the $\nu_e$ number density,
$F_{\nu_e}(\omega,\mathbf{v})$ with $\omega>0$ ($\omega<0$) is the $\nu_e$ ($\bar\nu_e$) spectral and angular distribution function, 
and $G(\omega,\mathbf{v})=\text{sgn}(\omega)[F_{\nu_e}(\omega,\mathbf{v})-F_{\nu_x}(\omega,\mathbf{v})]/n_{\nu_e}$.
The contraction $v^\rho(\mathbf{v}) v_\rho(\mathbf{v}')$ gives the factor $(1-\mathbf{v}\cdot\mathbf{v}')$ that is explicitly
included in the usual expression of $\mathsf{H}_{\nu\nu}(\mathbf{v})$. Here we emphasize the physical importance of
$\mathsf{J}^\rho$ and will discuss neutrino flavor evolution using its 12 components.

We focus on the so-called fast flavor oscillations \cite{sawyer2005speed,sawyer2009multiangle}. 
When the angular distribution $\int_{-\infty}^{+\infty}d\omega\, G(\omega,\mathbf{v})$ of the electron 
lepton number ($\nu$ELN) carried by a dense neutrino gas has a zero-crossing, i.e., 
switches from being positive to negative, an instability may be triggered, which could result in fast flavor 
conversion on length scales as short as $\sim\mathcal{O}(1\text{m})$ for conditions in neutron star mergers 
\cite{wu2017fast,wu2017imprints} and supernovae \cite{dasgupta2017fast,tamborra2014self,morinaga2020fast}.
Whereas the above flavor instability can be identified by a linear stability analysis
\cite{izaguirre2017fast,dasgupta2018simple,yi2019dispersion}, the eventual outcome of fast oscillations is
much harder to ascertain for realistic astrophysical environments. 
Apart from the uncertainties in modeling the neutrino emission in supernovae and neutron star mergers, 
numerical treatment of neutrino flavor evolution is further hampered by the complicated geometry and 
intrinsically dynamic nature of such environments. Consequently, studies of fast oscillations beyond 
the linear regime have been restricted to greatly simplified models so far. Specifically,
Eq.~(\ref{eq:pvw_vdrho_P}) was solved for artificial $\nu$ELN angular distributions by keeping 
the spatial derivative only in one \cite{martin2020dynamic,bhattacharyya2020late,bhattacharyya2020fast} 
or two directions \cite{shalgar2020neutrino}. Other studies dropped either the time
\cite{abbar2019fast} or spatial derivative \cite{johns2020neutrino,johns2020fast,shalgar2020dispelling}.

In this letter, we assume a homogeneous neutrino gas for which the spatial derivative can be ignored.
We find the stationary solutions and compare them to the results from evolution calculations. We also 
discuss improvement of these solutions and their use as estimates of the effects of
fast oscillations in astrophysical environments.

\section{Stationary solutions}
\label{sec:ss}
Under our assumption of homogeneity, Eq.~(\ref{eq:pvw_vdrho_P}) becomes
\begin{equation}
	\partial_t\mathsf{P}(\omega,\mathbf{v}) = \mathsf{H}(\omega,\mathbf{v}) \times \mathsf{P}(\omega,\mathbf{v}).
	\label{eq:ptP}
\end{equation}
The evolution of $\mathsf{P}(\omega,\mathbf{v})$ for the initial $\nu_e$ ($\omega>0$) differs from
that for the initial $\bar\nu_e$ ($\omega<0$) only through the vacuum term $\mathsf{H}_\text{vac}(\omega)$
in $\mathsf{H}(\omega,\mathbf{v})$.
For the dense astrophysical environments of interest to us, $\mathsf{H}_\text{vac}(\omega)$
can be ignored because the magnitude of $\omega$ is far less than that of $\lambda$ or $\mu$ associated with
$\mathsf{H}_\text{mat}$ or $\mathsf{H}_{\nu\nu}(\mathbf{v})$, respectively.
The effect of $\mathsf{H}_\text{vac}(\omega)$ is to initiate neutrino flavor mixing, which can be 
approximated by allowing $\mathsf{P}(\omega,\mathbf{v})$ to have small initial deviations from the pure flavor states. 
With this prescription, the evolution of $\mathsf{P}(\omega,\mathbf{v})$ no longer depends on $\omega$.
We write $\mathsf{P}(\omega,\mathbf{v})=\mathsf{P}(\mathbf{v})$ and solve the evolution equation
\begin{equation}
	\partial_t \mathsf{P}(\mathbf{v}) = \mathsf{H}(\mathbf{v}) \times \mathsf{P}(\mathbf{v}),
	\label{eq:pv_dt_P}
\end{equation}
where $\mathsf{H}(\mathbf{v}) = \lambda \hat{\mathsf{e}}_3 + v^\rho(\mathbf{v}) \mathsf{J}_\rho$,
$\mathsf{J}^\rho=\mu\int d\mathbf{v}'\, g(\mathbf{v}') {v}^\rho(\mathbf{v}') \mathsf{P}(\mathbf{v}')$, and
$g(\mathbf{v})=\int_{-\infty}^{+\infty}d\omega\, G(\omega,\mathbf{v})$ is the angular $\nu$ELN distribution.
From Eq.~(\ref{eq:pv_dt_P}), we obtain $\partial_t\mathsf{J}^t = \lambda \hat{\mathsf{e}}_3 \times \mathsf{J}^t$.
For the small initial deviations of $\mathsf{P}(\mathbf{v})$ from the pure flavor states,
$\mathsf{J}^t$ is essentially parallel to $\hat{\mathsf{e}}_3$ initially (at $t=t_0$). Therefore,
$\partial_t\mathsf{J}^t\approx 0$ and 
$\mathsf{J}^t\approx\mathsf{J}^t_{t_0}\approx\mu_{\rm eff}\hat{\mathsf{e}}_3$, where
$\mu_{\rm eff}=\mu\int d\mathbf{v}\, g(\mathbf{v})$ is specified by the $\nu{\rm ELN}$.

Based on the above discussion, we take 
\begin{equation}
\mathsf{H}(\mathbf{v}) = (\lambda+\mu_{\rm eff}) \hat{\mathsf{e}}_3 -(v^x\mathsf{J}^x+v^y\mathsf{J}^y+v^z\mathsf{J}^z).
\label{eq:Hv}
\end{equation}
The components of the polarization current $\mathsf{J}^x$, $\mathsf{J}^y$, and $\mathsf{J}^z$ are vectors 
in the flavor space. At any specific time, the range of $\mathsf{H}(\mathbf{v})$ is determined by these
vectors and the constraint $(v^x)^2+(v^y)^2+(v^z)^2=1$, and each $\mathsf{P}(\mathbf{v})$ precesses with the
corresponding angular velocity $\mathsf{H}(\mathbf{v})$. We seek stationary solutions for which
all polarization vectors collectively precess with the same angular velocity $\mathsf{\Omega}$ on average.
In the frame that co-rotates with these vectors, the evolution equation becomes
\begin{equation}
	\tilde{\partial}_t \mathsf{P}(\mathbf{v})=[\mathsf{H}(\mathbf{v})-\mathsf{\Omega}] \times \mathsf{P}(\mathbf{v})
	=\mathsf{H}'(\mathbf{v})\times \mathsf{P}(\mathbf{v}).
	\label{eq:pv_OmegaXP}
\end{equation}
We assume adiabatic evolution, for which the angle between $\mathsf{P}(\mathbf{v})$ and 
$\mathsf{H}'(\mathbf{v})=\mathsf{H}(\mathbf{v})-\mathsf{\Omega}$ stays fixed
(see similar approach in \cite{duan2007picture,wu2016physics}). 
Because of the constraint $\mathsf{J}^t\approx\mu_{\rm eff}\hat{\mathsf{e}}_3$, $\mathsf{\Omega}$ is approximately
parallel to $\hat{\mathsf{e}}_3$ for the stationary solutions. For the small initial deviations of $\mathsf{P}(\mathbf{v})$
from the pure flavor states, $\mathsf{J}^x$, $\mathsf{J}^y$, and $\mathsf{J}^z$ are also approximately parallel to 
$\hat{\mathsf{e}}_3$ at time $t_0$. So $\mathsf{P}(\mathbf{v})$ is parallel to 
$\mathsf{H}'(\mathbf{v})$ initially, and the adiabatic condition can be written as
\begin{equation}
	\mathsf{P}(\mathbf{v}) = \epsilon(\mathbf{v}) \hat{\mathsf{H}}'(\mathbf{v}),
	\label{eq:cf_alignment}
\end{equation}
where $\hat{\mathsf{H}}'(\mathbf{v})$ is the unit vector in the direction of $\mathsf{H}'(\mathbf{v})$, and
\begin{equation}
	\epsilon(\mathbf{v}) = \text{sgn} \left\{ \hat{\mathsf{e}}_3 \cdot \left[\mathsf{H}_{t_0}(\mathbf{v}) - \mathsf{\Omega} \right] \right\}
	=\text{sgn}\left[H'_{3,t_0}(\mathbf{v})\right].
	\label{eq:cf_epsilon1}
\end{equation}
The angular velocity $\mathsf{\Omega}$ and the polarization current components $\mathsf{J}^x$, $\mathsf{J}^y$, and $\mathsf{J}^z$
for the stationary solutions can be solved iteratively by using Eqs.~(\ref{eq:cf_alignment}) and (\ref{eq:cf_epsilon1}) in the definition of 
$\mathsf{J}^\rho$ and applying the constraint $\mathsf{J}^t\approx\mu_{\rm eff}\hat{\mathsf{e}}_3$. Clearly, it is most
convenient to carry out the above procedure in the co-rotating frame.

\section{Example solutions}
\label{sec:results}
\begin{figure}[ht!]
	\centering
		\includegraphics[width=0.35\textwidth]{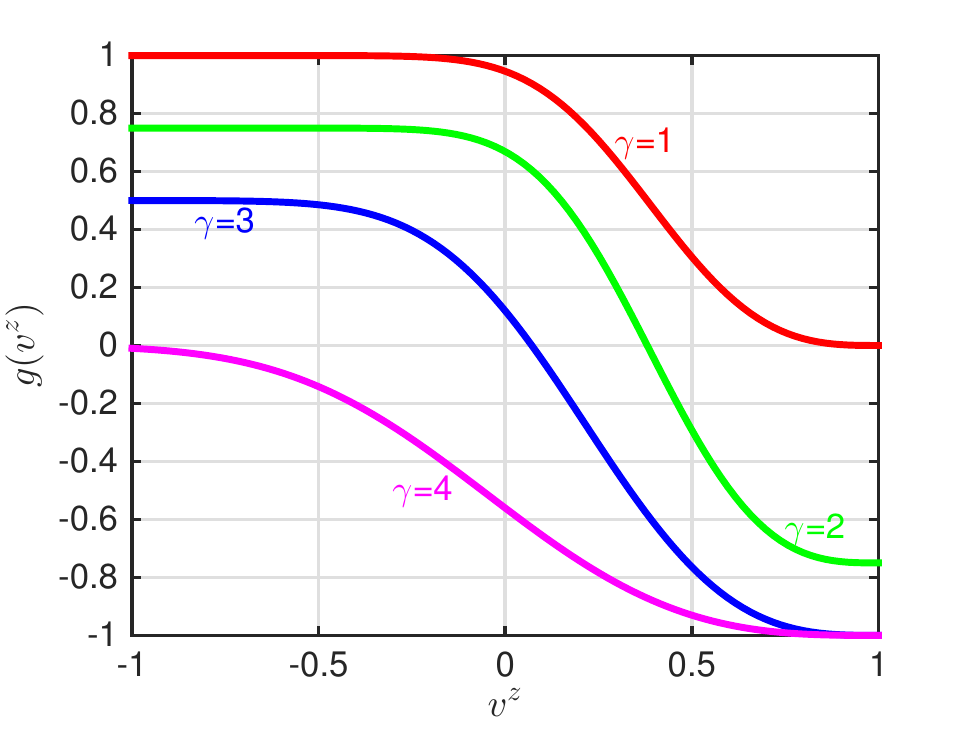}
	\caption{\label{fig:g_dist} The assumed $\nu$ELN distributions $g(v^z)$.
	The range of $\gamma=1$--4 allows any zero-crossing between $v^z=1$ and $-1$.}
\end{figure}

We now illustrate the stationary solutions with specific examples. Because the angular velocity $\mathsf{\Omega}$
is approximately parallel to the matter term $\lambda \hat{\mathsf{e}}_3$ in $\mathsf{H}$, $\lambda$ effectively
shifts the magnitude of $\mathsf{\Omega}$. We take $\lambda=0$ and $\mu=10^4/(4\pi)$~km$^{-1}$. Motivated by the 
conditions in supernovae \cite{abbar2020fast}, we assume azimuthally symmetric $\nu$ELN distributions 
$g(\mathbf{v})=g(v^z)$, where
\begin{align}
    g(v^z)
    = 
    \begin{cases}
        1.25-0.25\gamma -(0.5+0.5\gamma)e^{-\left[(1-v^z)/0.7\right]^3}, & 1 \leq \gamma \leq 2,\\
        1.25-0.25\gamma -1.5e^{-\left[(1-v^z)/(0.3+0.2\gamma)\right]^3}, & 2 < \gamma \leq 3,\\
        2-0.5\gamma -(3-0.5\gamma)e^{-\left[(1-v^z)/(0.3\gamma)\right]^3}, & 3 < \gamma \leq 4.
    \end{cases}
	 \label{eq:g_dist}
\end{align}
The range of $\gamma=1$--4 allows the above $g(v^z)$ to have any zero-crossing between
$v^z=1$ and $-1$ (see Fig.~\ref{fig:g_dist}). We focus on the cases of $\gamma=2$ and 3 with a crossing
at $v^z\approx0.38$ and 0.07, respectively. 

\begin{figure}[ht!]
\centering
\subfigure{\includegraphics[width=0.22\textwidth]{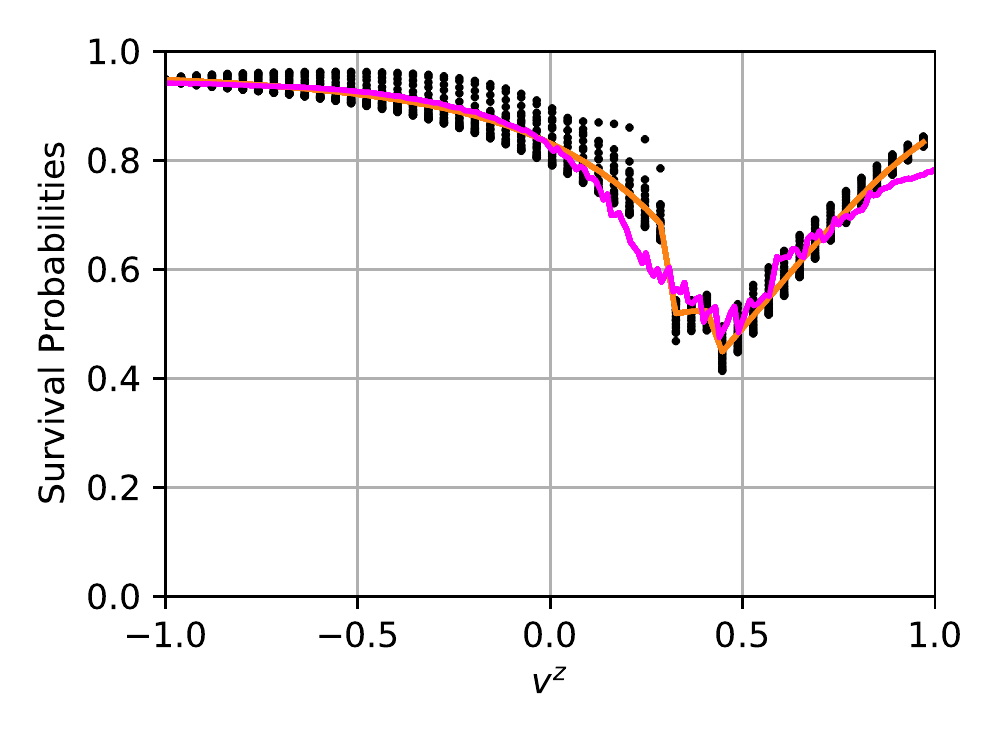}
\llap{\parbox[b]{1.3in}{\small (a)\\\rule{0ex}{0.25in}}} }
\subfigure{\includegraphics[width=0.22\textwidth]{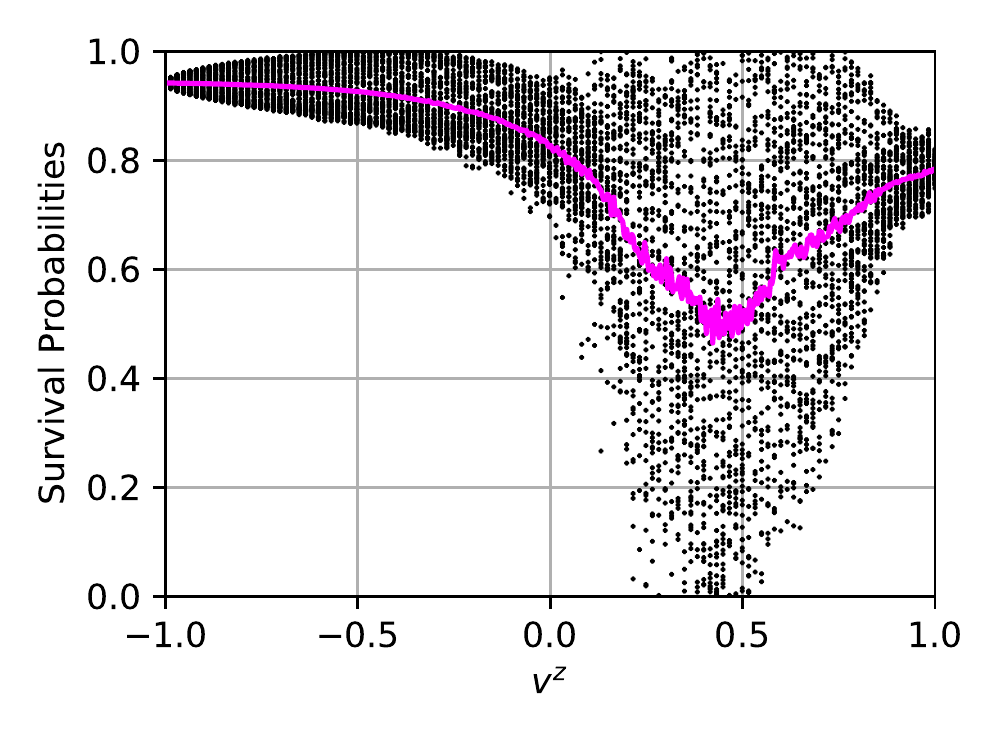}
\llap{\parbox[b]{1.3in}{\small (b)\\\rule{0ex}{0.25in}}} }\\
\subfigure{\includegraphics[width=0.22\textwidth]{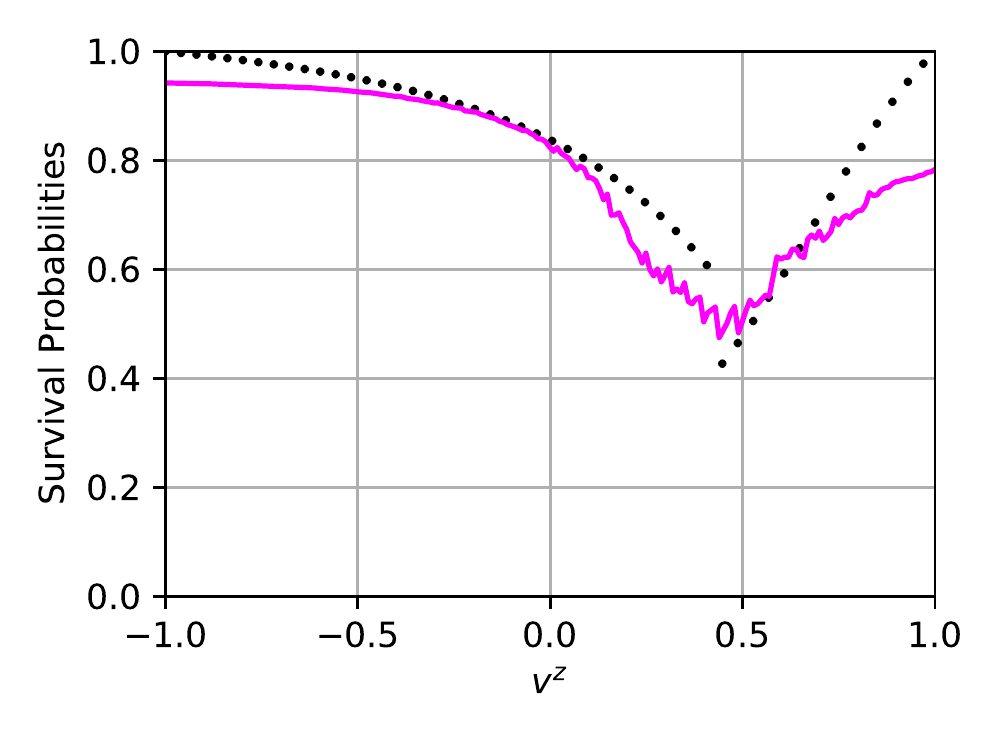}
\llap{\parbox[b]{1.3in}{\small (c)\\\rule{0ex}{0.25in}}} }
\subfigure{\includegraphics[width=0.22\textwidth]{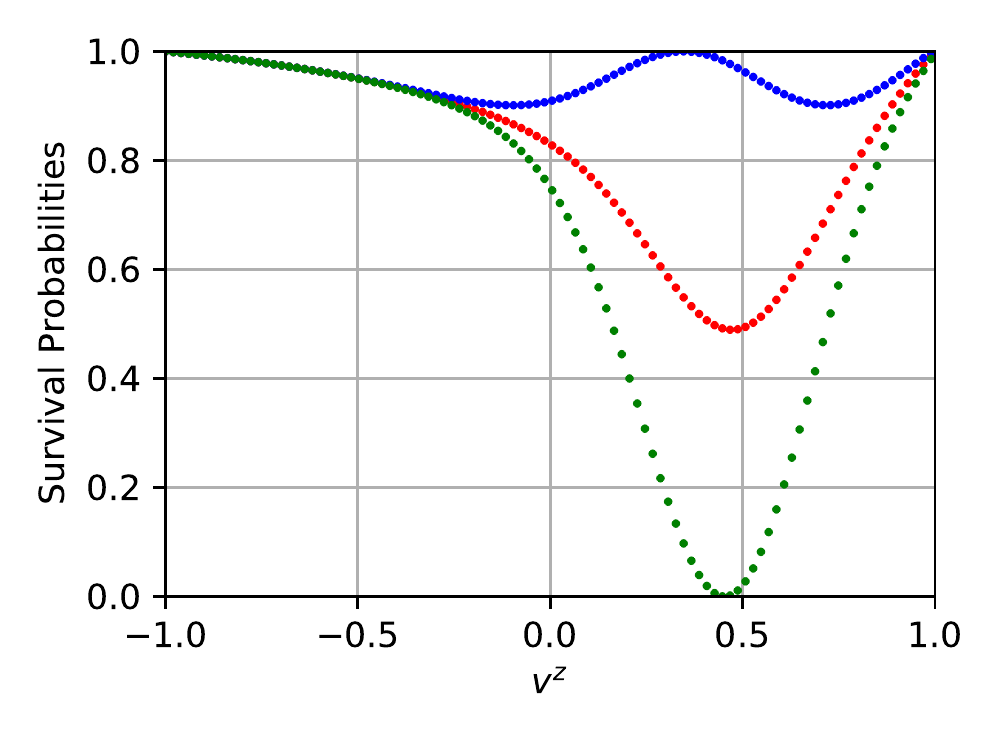}
\llap{\parbox[b]{1.3in}{\small (d)\\\rule{0ex}{0.25in}}} }\\
\subfigure{\includegraphics[width=0.22\textwidth]{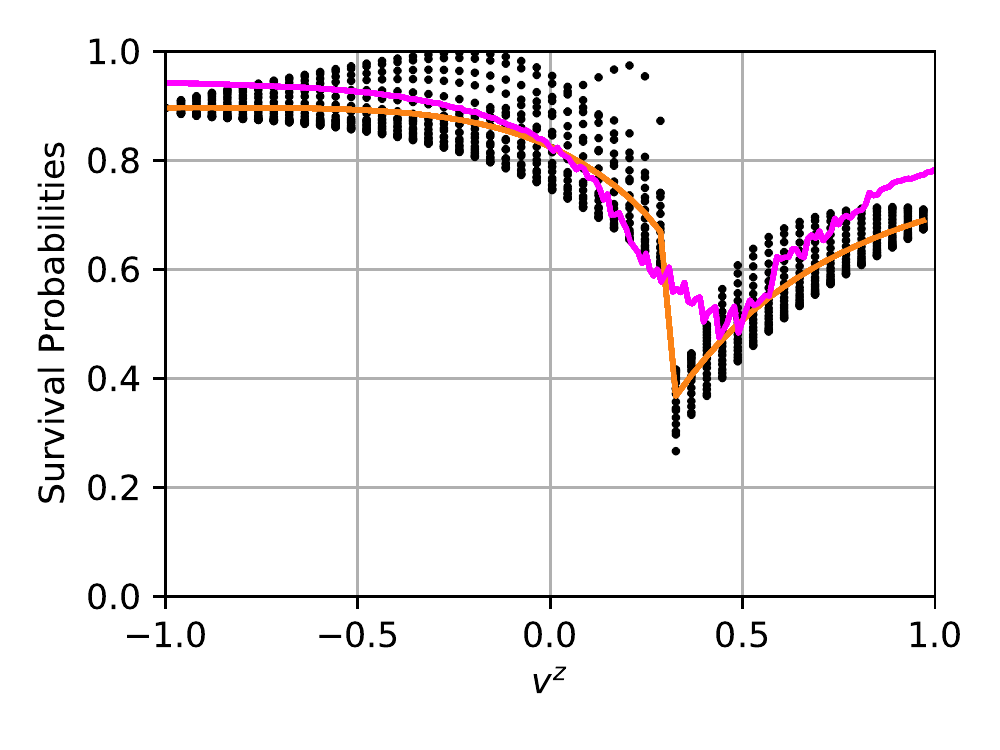}
\llap{\parbox[b]{1.3in}{\small (e)\\\rule{0ex}{0.25in}}} }
\subfigure{\includegraphics[width=0.22\textwidth]{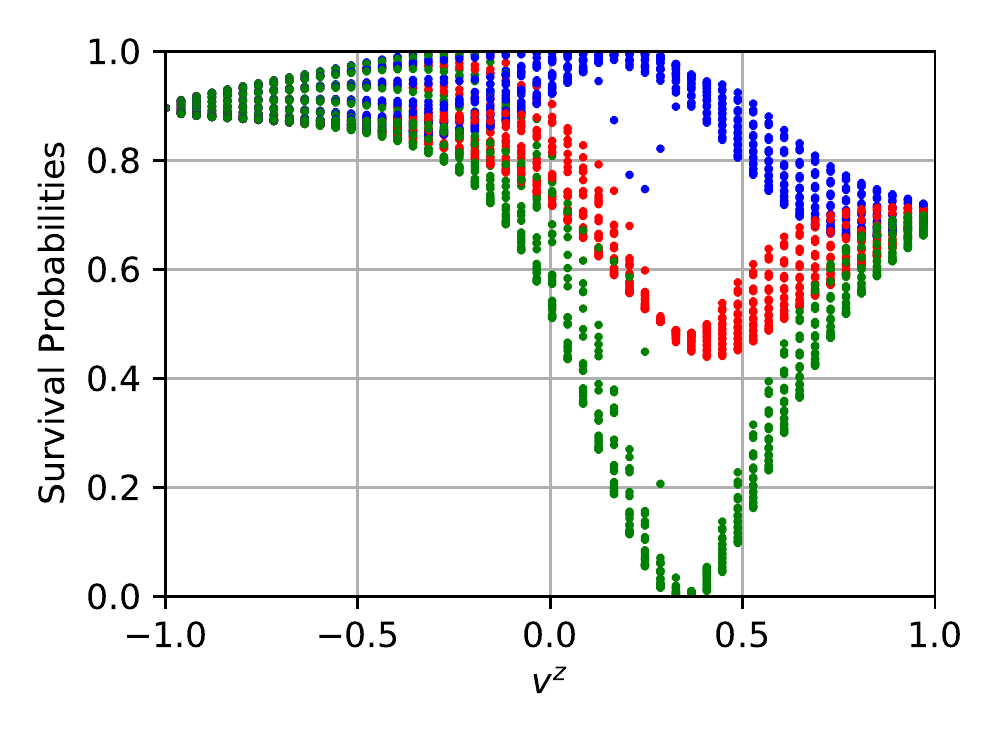}
\llap{\parbox[b]{1.3in}{\small (f)\\\rule{0ex}{0.25in}}} }
	\caption{\label{fig:P_comparison_2} 
	Comparison of the survival probabilities from the stationary
	solutions with the numerical results for $\gamma=2$. 
	Symbols at the same $v^z$ are for different azimuthal angles.
	In panel (a), the symbols are the averages of the results for the stationary solutions
	[shown in panels (c) and (e)] and the orange curve is obtained by averaging these symbols 
	over the azimuthal angle. In panel (b), the symbols are the numerical results and 
	the magenta curve is obtained by averaging these symbols over the azimuthal angle.
	The same magenta curve is also shown in panels (a), (c), and (e) for comparison.
	In panel (c), the symbols show the stationary solution of type IIIa. In panel (e), the symbols
	show the stationary solution of type IVa and the orange curve is obtained by averaging
	these symbols over the azimuthal angle. In panels (d) and (f), the symbols show
	the survival probabilities (red: mean, blue and green: limits) including the approximate effect 
	of nonadiabatic evolution for the stationary solutions of types IIIa and IVa, respectively.}
\end{figure}

\begin{figure}[ht!]
\centering
\subfigure{\includegraphics[width=0.22\textwidth]{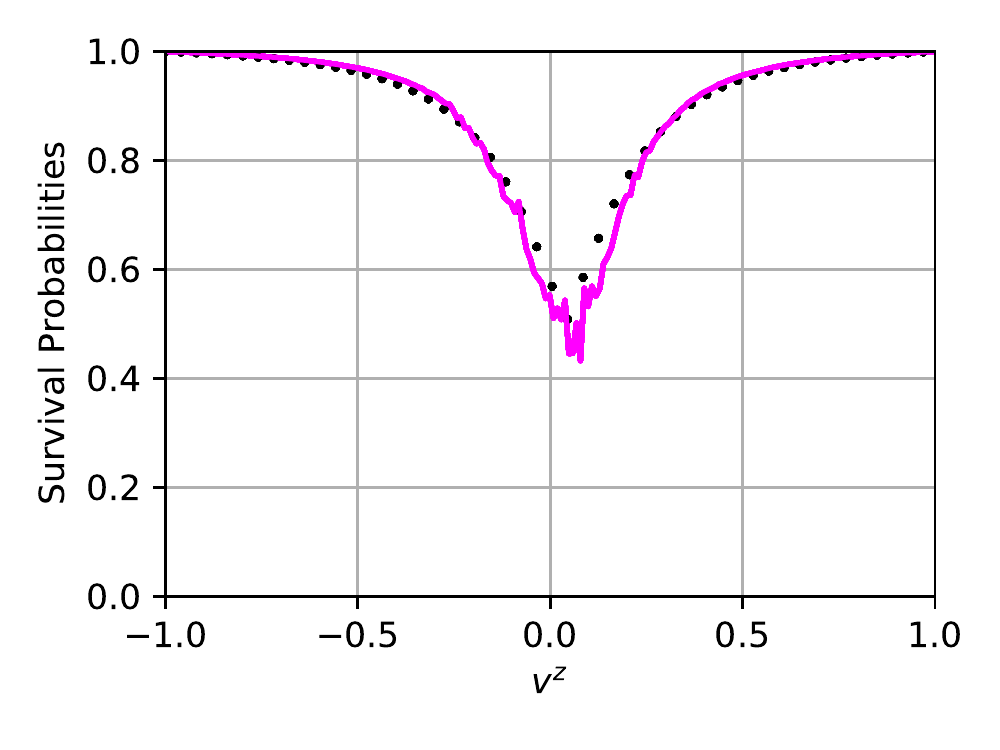}
\llap{\parbox[b]{1.3in}{\small (a)\\\rule{0ex}{0.25in}}} }
\subfigure{\includegraphics[width=0.22\textwidth]{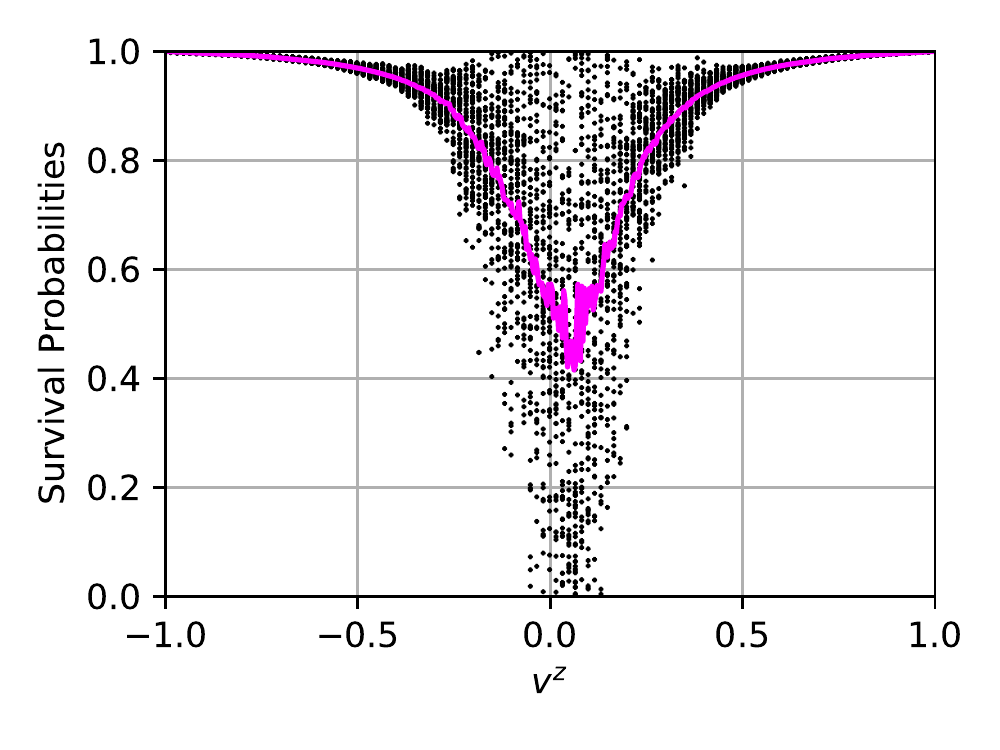}
\llap{\parbox[b]{1.3in}{\small (b)\\\rule{0ex}{0.25in}}} }\\
\subfigure{\includegraphics[width=0.22\textwidth]{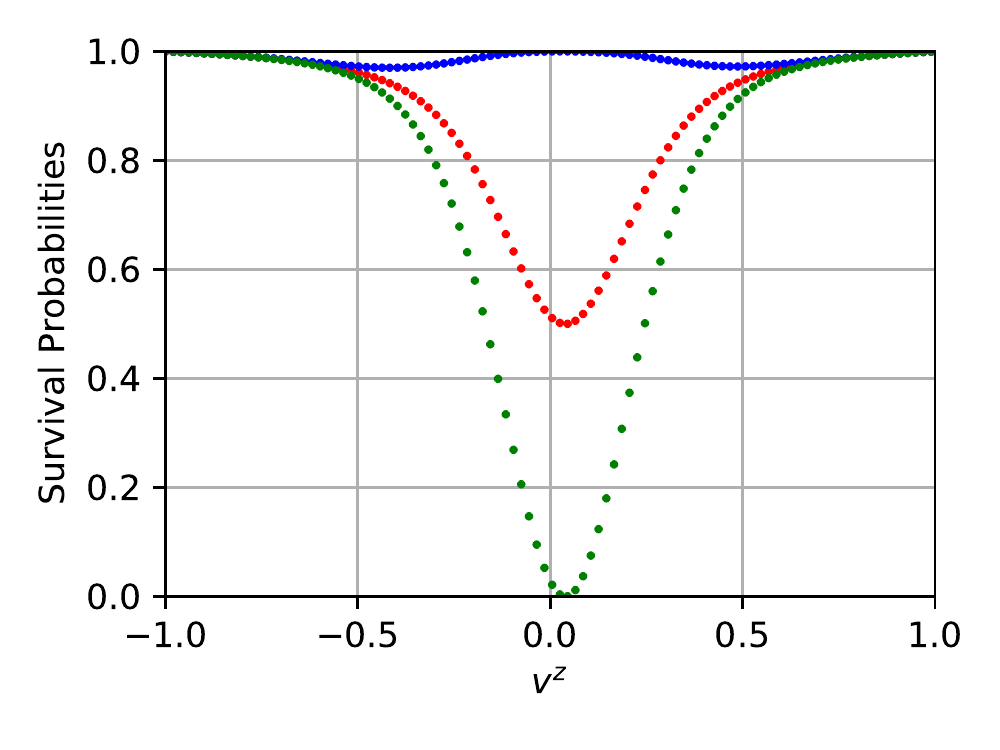}
\llap{\parbox[b]{1.3in}{\small (c)\\\rule{0ex}{0.25in}}} }
\subfigure{\includegraphics[width=0.22\textwidth]{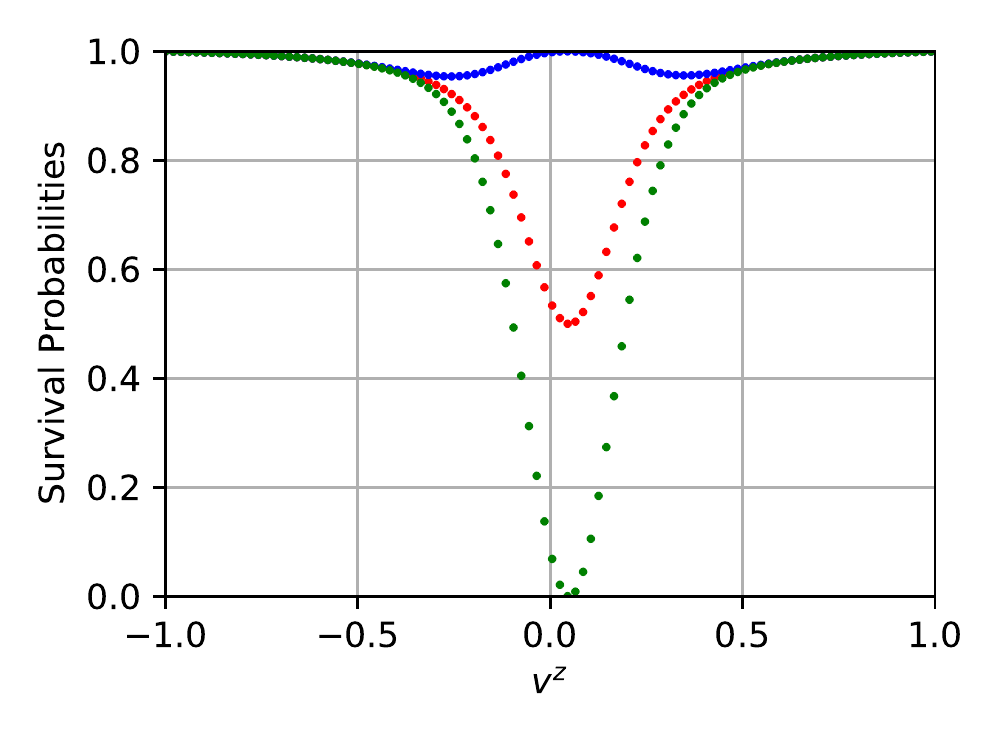}
\llap{\parbox[b]{1.3in}{\small (d)\\\rule{0ex}{0.25in}}} }
	\caption{\label{fig:P_comparison_3} Similar to Fig.~\ref{fig:P_comparison_2}, but for
	$\gamma=3$. In panel (a), the symbols show the stationary solution of type IIIa and
	the magenta curve shows the survival probabilities averaged over the azimuthal angle
	for the numerical results in panel (b). In panel (c), the symbols show
	the survival probabilities (red: mean, blue and green: limits) including the approximate effect 
	of nonadiabatic evolution based on the stationary solution in panel (a). Panel (d) is the same
	as panel (c), but for a new stationary solution of type IIIa$'$ characterized by a new
	ansatz for the neutrino polarization vectors.}
\end{figure}

We first solve Eq.~(\ref{eq:pv_dt_P}) numerically using 600 bins for $v^z$ and 320 bins for the azimuthal 
angle of $\mathbf{v}$. We have checked that convergence is achieved for this angular resolution.
All polarization vectors $\mathsf{P}(\mathbf{v})$ are assigned random deviations
between $-10^{-3}$ and $10^{-3}$ from $P_3(\mathbf{v})=1$ at $t=t_0$. Their evolution is followed up to
$t-t_0=2$~km, when an approximately stationary state has been reached (see \ref{sec:convergence}).
The results on the survival probability 
$(P_3+1)/2$ for $\gamma=2$ and 3 are shown in Figs.~\ref{fig:P_comparison_2}(b) and \ref{fig:P_comparison_3}(b),
respectively. Because the azimuthal symmetry is broken by the initial conditions adopted to approximate 
the effects of the vacuum term in $\mathsf{H}(\omega,\mathbf{v})$, these results depend on the azimuthal 
angle of $\mathbf{v}$ for each bin of $v^z$. The corresponding values of $\Omega$, $\mathsf{J}^x$, 
$\mathsf{J}^y$, and $\mathsf{J}^z$ are given in Table~\ref{tab:solutions}.

\begin{table}[ht!]
\caption{\label{tab:solutions} Characteristics of stationary solutions.}
    \begin{center}
	\begin{tabular}{llcccccccccc}\hline\hline
	$\gamma$ & results & $\Omega$ & $J^x_\perp$ & $J^y_\perp$ & $J^z_\perp$ & $J^x_3$ & $J^z_3$ \\\hline
	2 & numerical & 3750 & 800 & 1200 & 1600 & 0 & $-2000$ \\	
	2 & IIIa & 4089 & 1392 & 1392 & 0 & 0 & $-2446$ \\
	2 & IVa  & 3705 & 462 & 154 & 2082 & 190 & $-1762$ \\
	3 & numerical & $-870$ & 650 & 650 & 0 & 0 & $-3200$ \\
	3 & IIIa & $-900$ & 799 & 799 & 0 & 0 & $-3108$ \\
	3 & IIIa$'$ & $-867$ & 622 & 622 & 0 & 0 & $-3161$\\\hline
	\end{tabular}
    \end{center}
\end{table}

We next calculate the stationary solutions as presented in Section~\ref{sec:ss}. For all practical
purposes, we can assume pure flavor states to obtain $\mathsf{J}^t=\mu_{\rm eff}\hat{\mathsf{e}}_3$, 
$\mathsf{J}^x=\mathsf{J}^y=0$, and $\mathsf{J}^z\parallel\hat{\mathsf{e}}_3$ at $t=t_0$, which gives
$\mathsf{H}_{t_0}(\mathbf{v})\parallel\hat{\mathsf{e}}_3$. Specifically, the initial conditions are 
$J^t_3=2812$~km$^{-1}$, $J^z_3=-3030$~km$^{-1}$ for $\gamma=2$ and $J^t_3=-1028$~km$^{-1}$, 
$J^z_3=-3286$~km$^{-1}$ for $\gamma=3$. Conservation of $\mathsf{J}^t$ requires 
$\mathsf{\Omega}=\Omega\hat{\mathsf{e}}_3$.
Due to the azimuthal symmetry of $g(v^z)$ around the
$z$ axis in the Euclidean space, we can specify the $x$ and $y$ axes by setting $J^y_3=0$ for the
stationary solutions. In addition, the rotational symmetry around $\hat{\mathsf{e}}_3$ in the flavor 
space allows us to specify the directions $\hat{\mathsf{e}}_1$ and $\hat{\mathsf{e}}_2$ of the 
corotating frame by setting $J^x_2=0$. The remaining components of the neutrino polarization current 
$J^x_1$, $J^y_1$, $J^z_1$, $J^y_2$, $J^z_2$, $J^x_3$, and $J^z_3$ for the stationary solutions
can be solved from the iterative procedure in Section~\ref{sec:ss}.

We find that the stationary solutions can be classified into the following types, with
the subscript $\perp$ denoting either $\hat{\mathsf{e}}_1$ or $\hat{\mathsf{e}}_2$:
(I) $J^{x,y,z}_\perp\sim 0$, which represents the initial configuration, and therefore, is trivial,
(II) $J^{x,y}_\perp\sim 0$ and $J^z_\perp\sim{\cal{O}}(\mu)$, 
(III) $J^{x,y}_\perp\sim{\cal{O}}(\mu)$ and $J^z_\perp\sim0$, and
(IV) $J^{x,y}_\perp\sim{\cal{O}}(\mu)$ and $J^z_\perp\sim{\cal{O}}(\mu)$.
Type III is further divided into two subtypes: (IIIa) $\mathsf{J}^x_\perp\perp\mathsf{J}^y_\perp$
($J^x_1\neq 0$, $J^y_1=0$, $J^y_2\neq 0$), and (IIIb) $\mathsf{J}^x_\perp\parallel\mathsf{J}^y_\perp$ 
($J^x_1\neq 0$, $J^y_1\neq 0$, $J^y_2=0$). Similarly, type IV is further divided into three subtypes: (IVa)
$\mathsf{J}^x_\perp\perp\mathsf{J}^y_\perp$ and $\mathsf{J}^x_\perp\parallel\mathsf{J}^z_\perp$
($J^x_1\neq 0$, $J^y_1=0$, $J^z_1\neq0$, $J^y_2\neq 0$, $J^z_2=0$),
(IVb) $\mathsf{J}^x_\perp\parallel\mathsf{J}^y_\perp$ and $\mathsf{J}^y_\perp\perp\mathsf{J}^z_\perp$
($J^x_1\neq 0$, $J^y_1\neq 0$, $J^z_1=J^y_2=0$, $J^z_2\neq 0$),
and (IVc) $\mathsf{J}^x_\perp\parallel\mathsf{J}^y_\perp\parallel\mathsf{J}^z_\perp$
($J^x_1\neq 0$, $J^y_1\neq 0$, $J^z_1\neq 0$, $J^y_2=J^z_2=0$). While the above classification
reflects specific relations among the neutrino polarization current components, such relations are not
used to find the stationary solutions but are observed after the solutions are obtained.
Note that large magnitudes of $J^{x,y,z}_\perp$ correspond to substantial overall flavor 
transformation with large magnitudes of $P_\perp$ for wide ranges of $\mathbf{v}$.

Because the procedure outlined in Section~\ref{sec:ss} involves solving integral equations, 
the results are not unique in general
and are only candidate stationary solutions. We perturb the polarization vectors $\mathsf{P}(\mathbf{v})$ of 
a candidate solution with random deviations of magnitude up to $10^{-3}$ and evolve them with 
Eq.~(\ref{eq:pv_dt_P}) until the stability of the solution can be established or otherwise.
The stable ones are selected as true stationary solutions. Note that the results may not be unique
even after the above stability test. 

For $\gamma=2$, we find seven candidate stationary solutions, two of which are stable (types IIIa and
IVa). In contrast, only one of the five candidate solutions are stable (type IIIa) for $\gamma=3$
(see \ref{sec:convergence}).
The survival probabilities $(P_3+1)/2$ corresponding to the true stationary solutions are shown in 
Figs.~\ref{fig:P_comparison_2} and \ref{fig:P_comparison_3} for comparison with the numerical results.
The corresponding values of $\Omega$, $\mathsf{J}^x$, $\mathsf{J}^y$, and $\mathsf{J}^z$ are given in 
Table~\ref{tab:solutions}. For both $\gamma=2$ and 3, the stationary solution of type IIIa has
$J^x_3=J^y_3=J^z_\perp=0$ and $\mathsf{J}^x_\perp\perp\mathsf{J}^y_\perp$ with $J^x_\perp=J^y_\perp$.
Consequently, $\hat{H}'_3$ and hence $P_3$ are independent of the azimuthal angle [see Eqs.~(\ref{eq:Hv})
and (\ref{eq:cf_alignment})] with the corresponding survival probabilities exhibiting azimuthal symmetry.
In contrast, for $\gamma=2$, the stationary solution of type IVa has $J^x_3\neq 0$ and $J^{x,y,z}_\perp\neq 0$.
Therefore, the corresponding survival 
probabilities depend on the azimuthal angle.

The $\nu$ELN distribution $g(v^z)$ crosses zero at $v^z\approx0.38$ and 0.07 
for $\gamma=2$ and 3, respectively. It can be seen from Figs.~\ref{fig:P_comparison_2} and \ref{fig:P_comparison_3}
that except for $v^z$ near the zero-crossings, the survival probabilities given by the stationary solutions 
show the same trends as the numerical results. For more detailed comparisons, we calculate the average
survival probability for each bin of $v^z$ by averaging the numerical results over the azimuthal angle.
Figure~\ref{fig:P_comparison_3}(a) shows that the stationary solution for $\gamma=3$ describes the average
survival probabilities very well away from the zero-crossing. Because there are two types of stationary
solutions for $\gamma=2$, we calculate the corresponding average survival probability for each bin of $v^z$ by
weighing each type of solution equally and then averaging the results over the azimuthal angle. 
Figure~\ref{fig:P_comparison_2}(a) shows that these average survival probabilities are also in good agreement 
with the numerical results away from the zero-crossing.

The large deviations of the survival probabilities for the stationary solutions from the numerical results for $v^z$
near the zero-crossings are caused by the breakdown of adiabatic evolution assumed in deriving these solutions.
The corresponding neutrinos have small values of 
$|\mathsf{H}_{t_0}'(\mathbf{v})|=|H_{3,t_0}'(\mathbf{v})|\ll\mu$ that result in nonadiabatic flavor evolution.
The effect of this nonadiabaticity can be approximated by including a component of $\mathsf{P}(\mathbf{v})$
that is perpendicular to and rapidly rotating around $\mathsf{H}'(\mathbf{v})$ in addition to the component aligned
with the latter and given by
\begin{equation}
    \mathsf{P}_{\rm align}(\mathbf{v})\approx\epsilon_{\rm eff}(\mathbf{v})\hat{\mathsf{H}}'(\mathbf{v}).
    \label{eq:Palign}
\end{equation}
In the above equation, $\hat{\mathsf{H}}'(\mathbf{v})$ is taken from the stationary solution and 
$\epsilon_{\rm eff}(\mathbf{v})$ is taken as the error function
\begin{equation}
    \epsilon_{\rm eff}(\mathbf{v})=\text{erf}\left[H_{3,t_0}'(\mathbf{v})/\mu\right],
    \label{eq:cf_epsilon2}
\end{equation}
which approaches $\epsilon(\mathbf{v})=\pm1$ in Eq.~(\ref{eq:cf_epsilon1}) for adiabatic evolution with 
$|H'_{3,t_0}/\mu|\gg1$. Because the magnitude of $\mathsf{P}(\mathbf{v})$ is conserved during the evolution,
its component perpendicular to $\mathsf{H}'(\mathbf{v})$ should have a magnitude of 
$\sqrt{1-[\epsilon_{\rm eff}(\mathbf{v})]^2}$\,. Due to the rapid rotation of this component around
$\mathsf{H}'(\mathbf{v})$, the effective $P_3$ oscillates between the limits
\begin{equation}
    P_{3,{\rm lim}}(\mathbf{v})=\bar P_3(\mathbf{v})\pm
    \sqrt{1-[\epsilon_{\rm eff}(\mathbf{v})]^2}\sqrt{1-[\hat{H}'_3(\mathbf{v})]^2}\,,
    \label{eq:Prange}
\end{equation}
where
\begin{equation}
    \bar P_3(\mathbf{v})=P_{3,{\rm align}}(\mathbf{v})\approx\epsilon_{\rm eff}(\mathbf{v})\hat{H}'_3(\mathbf{v})
    \label{eq:Pmean}
\end{equation}
is the mean value due to the component aligned with $\mathsf{H}'(\mathbf{v})$. 
The corresponding survival probabilities are shown as the blue, green (limits), and
red (mean) symbols in Figs.~\ref{fig:P_comparison_2}(d), \ref{fig:P_comparison_2}(f) 
for $\gamma=2$ and in Fig.~\ref{fig:P_comparison_3}(c) for $\gamma=3$. It can be seen that these results
describe both the trends and the range of the numerical results rather well.

Because the rapidly rotating component of $\mathsf{P}(\mathbf{v})$ perpendicular to $\mathsf{H}'(\mathbf{v})$
is essentially averaged out, only $\mathsf{P}_{\rm align}(\mathbf{v})$ is effectively used to find the stationary 
solutions. We can choose $\epsilon_{\rm eff}(\mathbf{v})$ and use Eq.~(\ref{eq:Palign}) to replace 
Eq.~(\ref{eq:cf_epsilon1}) in the procedure to find these solutions. As an example, we choose
\begin{equation}
    \epsilon'_\text{eff}(\mathbf{v}) = \text{erf}\left[1.5H_{3,t_0}'(\mathbf{v})/\mu\right]
	\label{eq:cf_epsilon3}
\end{equation}
for $\gamma=3$ and obtain a new stationary solution, which is also of type IIIa (denoted as IIIa$'$).
The corresponding survival probabilities (mean and limits) are calculated in the same way as for
Fig.~\ref{fig:P_comparison_3}(c) and shown in Fig.~\ref{fig:P_comparison_3}(d). 
The corresponding values of $\Omega$, $\mathsf{J}^x$, $\mathsf{J}^y$, and $\mathsf{J}^z$ are given in 
Table~\ref{tab:solutions}. It can be seen that the new stationary solution is much closer to the numerical results.
While the effect of nonadiabatic evolution can be well approximated by choosing an appropriate 
$\epsilon'_\text{eff}(\mathbf{v})$ as in the above example, we note, however, that our original procedure
to find the stationary solutions assuming adiabatic evolution is more straightforward and can already produce
average survival probabilities to good approximation.

As discussed above, nonadiabatic evolution appears to be associated with the zero-crossing of the
$\nu$ELN distribution $g(v^z)$. On the other hand, if $g(v^z)$ has no zero-crossing as for
$\gamma=1$ or 4 (see Fig.~\ref{fig:g_dist}), the initial values of $P_3(\mathbf{v})=1$ make $J^t_3$ 
reach the most positive or negative value for $\gamma=1$ or 4, respectively. Conservation of $J^t_3$ then ensures
$P_3(\mathbf{v})=1$ subsequently, and therefore, there is no flavor evolution at all \cite{izaguirre2017fast}.

\section{Discussion and conclusions}
\label{sec:conclusions}
We have presented a method to find the stationary solutions for fast flavor oscillations of a homogeneous
dense neutrino gas whose angular $\nu$ELN distribution has a zero-crossing. These solutions correspond 
to collective precession of all neutrino polarization vectors around a fixed axis in the flavor space 
on average, and are conveniently studied in the co-rotating frame. We have shown that these solutions 
can account for the numerical results of explicit evolution calculations, and that even with the simplest
assumption of adiabatic evolution, they can provide the average survival probabilities to good approximation. 
These solutions can be further improved by including the effect of nonadiabatic evolution, which in turn,
can be approximated by choosing an appropriate ansatz for the alignment of the polarization vectors with 
the effective Hamiltonian in the co-rotating frame. While we have focused on specific $\nu$ELN distributions 
for our examples, we show in \ref{sec:app_other_spectra} that our method applies to other $\nu$ELN 
distributions as well.

Besides shedding light on the nonlinear regime of fast oscillations, the stationary solutions
discussed here provide physically motivated estimates of the average survival probabilities beyond
the simple limit of complete flavor equilibration. Because these solutions can be efficiently calculated,
they may be incorporated in simulations of dynamical astrophysical environments such as supernovae 
and neutron star mergers, for which the computational resources must be almost exclusively devoted to 
hydrodynamics and regular neutrino transport by necessity. We note that our method assumes a 
homogeneous neutrino gas but the environments where fast oscillations may occur are most likely
inhomogeneous. Consequently, the stationary solutions discussed here may only provide a crude yet
efficient way to explore the effects of fast oscillations on neutrino-related processes in supernovae and 
neutron star mergers. The general problem of flavor evolution of an inhomogeneous dense neutrino gas is 
beyond our scope here and requires further dedicated studies.

\section*{Acknowledgements}
This work was supported in part by the US Department of Energy [DE-FG02-87ER40328].
We thank the anonymous referee for constructive criticisms and helpful suggestions.
Z. X. was partly supported by the European Research Council (ERC) under the European Union's 
Horizon 2020 research and innovation programme (ERC Advanced Grant KILONOVA No.~885281).
Some calculations were carried out at the Minnesota Supercomputing Institute.

\appendix

\section{Numerical tests}
\label{sec:convergence}
\begin{figure}[ht!]
\centering
\subfigure{\includegraphics[width=0.22\textwidth]{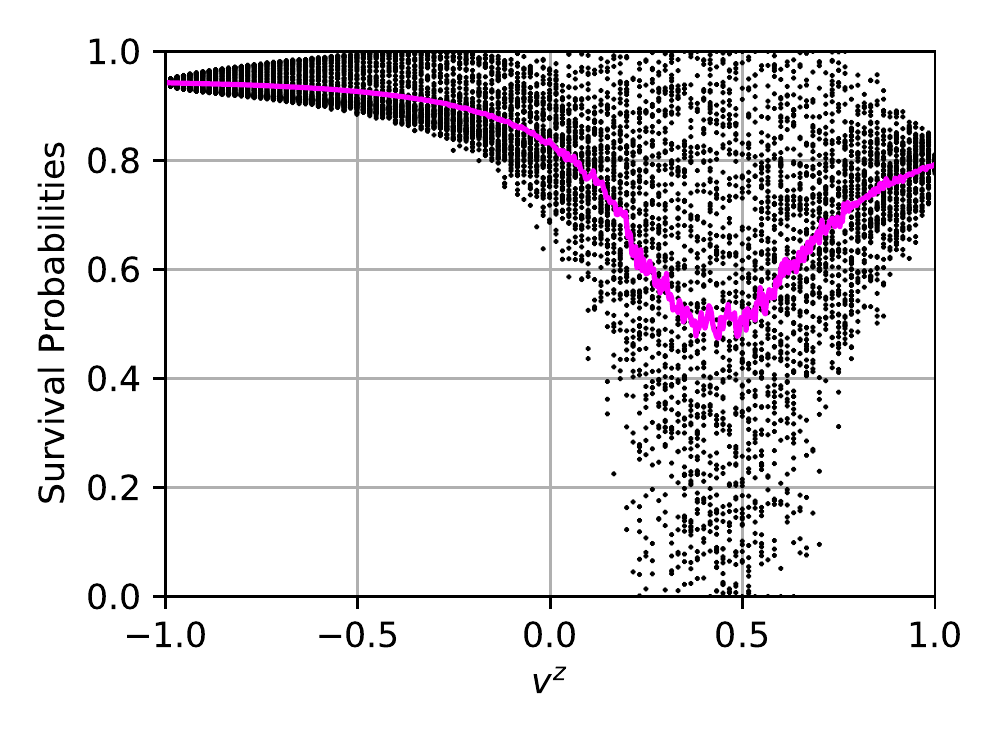}
\llap{\parbox[b]{1.3in}{\small (a)\\\rule{0ex}{0.25in}}} }
\subfigure{\includegraphics[width=0.22\textwidth]{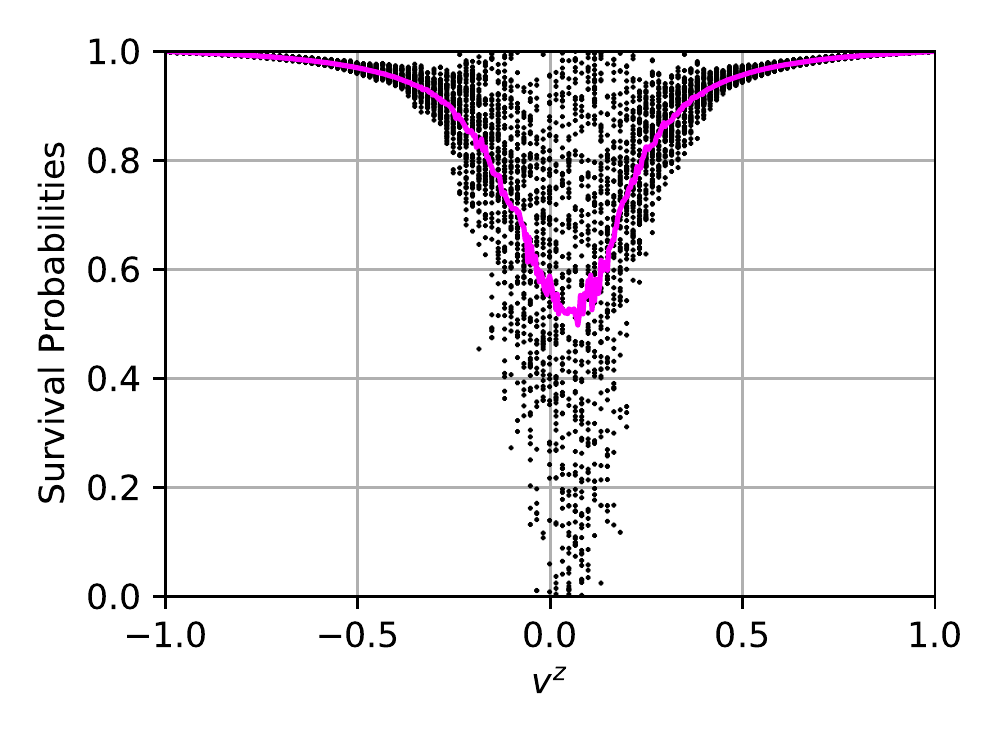}
\llap{\parbox[b]{1.3in}{\small (d)\\\rule{0ex}{0.25in}}} }\\
\subfigure{\includegraphics[width=0.22\textwidth]{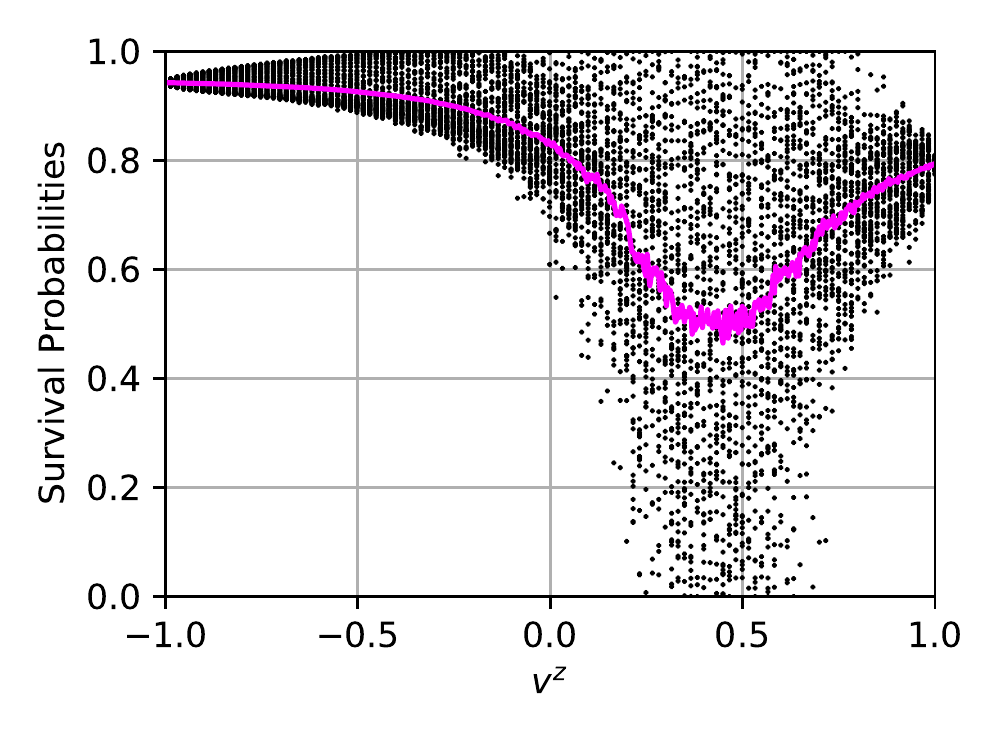}
\llap{\parbox[b]{1.3in}{\small (b)\\\rule{0ex}{0.25in}}} }
\subfigure{\includegraphics[width=0.22\textwidth]{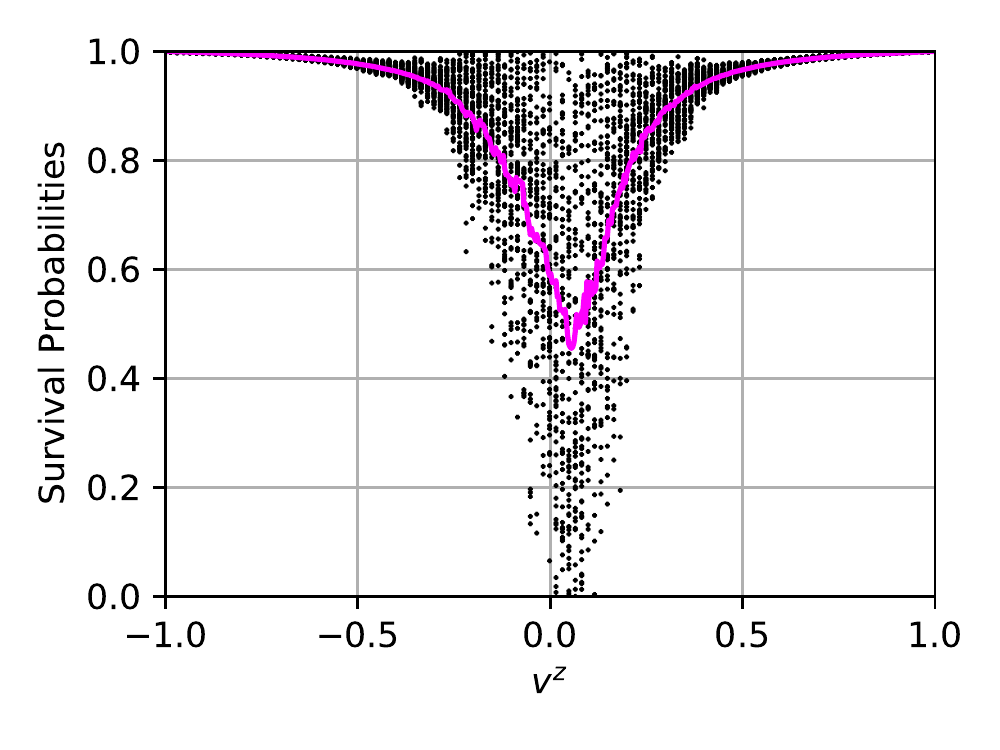}
\llap{\parbox[b]{1.3in}{\small (e)\\\rule{0ex}{0.25in}}} }\\
\subfigure{\includegraphics[width=0.22\textwidth]{2I_t2}
\llap{\parbox[b]{1.3in}{\small (c)\\\rule{0ex}{0.25in}}} }
\subfigure{\includegraphics[width=0.22\textwidth]{3I_t2}
\llap{\parbox[b]{1.3in}{\small (f)\\\rule{0ex}{0.25in}}} }
	\caption{\label{fig:convergence} Snapshots of survival probabilities.
	Panels (a), (b), and (c) [(d), (e), and (f)] are for $t=1.6$, 1.8, and 2~km,
	respectively, for the $\nu$ELN distribution with $\gamma=2$ (3).}
\end{figure}

Figure~\ref{fig:convergence} shows the snapshots of our explicit evolution calculations
at $t=1.6$, 1.8, and 2 $\mathrm{km}$ for the assumed $\nu$ELN distributions with
$\gamma=2$ and 3. It can be seen that an approximately stationary state has been reached
by $t=2$~km in both cases.

\begin{figure}[ht!]
\centering
\includegraphics[width=0.48\textwidth]{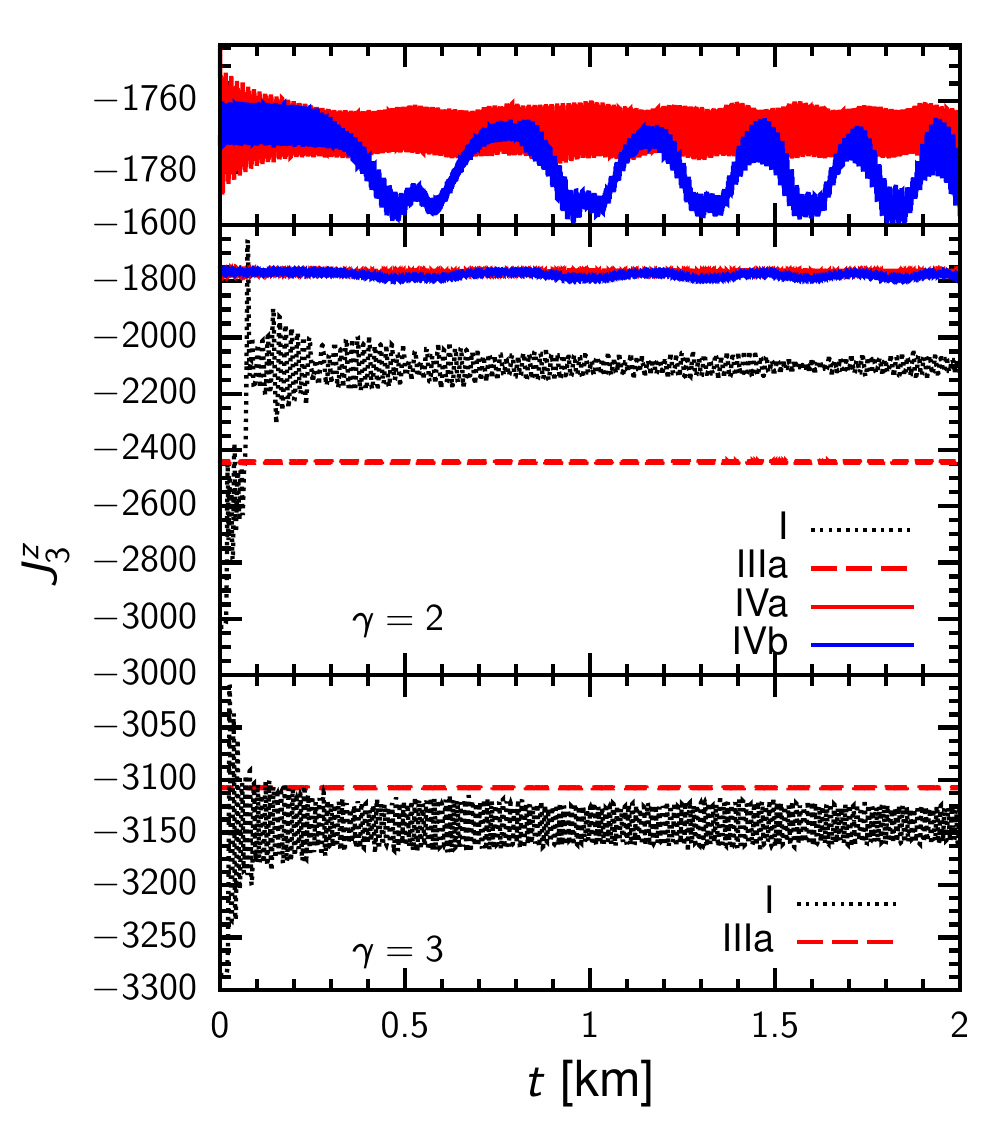}
	\caption{\label{fig:convergence2} Evolution of $J^z_3$ for some candidate
	stationary solutions after they are perturbed. For the $\nu$ELN distribution with $\gamma=2$, 
	the difference between types IVa and IVb is highlighted in the top panel.}
\end{figure}

Figure~\ref{fig:convergence2} shows the evolution of $J^z_3$ for some candidate stationary 
solutions after they are perturbed. Note that the results for type I are calculated from
the initial configurations and become approximately constant at late times as expected of 
the approximately stationary states shown in Fig.~\ref{fig:convergence}.
For $\gamma=2$, only types IIIa and IVa are the true stationary solutions. Type IVb is 
quasi-stable for $t<0.3$~km but deviates from the initial state subsequently. Therefore, 
it is not a true stationary solution. Note that the average value of $J^z_3$ for types IIIa
and IVa is close to the late-time value of $J^z_3$ for type I.
For $\gamma=3$, only type IIIa is the true stationary solution.

\section{Other $\nu$ELN distributions}
\label{sec:app_other_spectra}
We have repeated the calculations for two more $\nu$ELN distributions taken from 
\cite{shalgar2020dispelling} (case A) and \cite{yi2019dispersion} (case B):
\begin{align}
    g_A(v^z) & = -\exp \left[-\frac{(\cos^{-1} v^z)^2}{2} \right]+0.5,\\
    g_B(v^z) & = 10  \exp\left[-\frac{(1-v^z)^2}{0.72} \right] - 10.66  \exp\left[-\frac{(1-v^z)^2}{0.562} \right].
\end{align}
The zero-crossing occurs at $v^z\approx 0.38$ and 0.6 for $g_A(v^z)$ and $g_B(v^z)$, respectively.

\begin{figure}[ht!]
\centering
\subfigure{\includegraphics[width=0.22\textwidth]{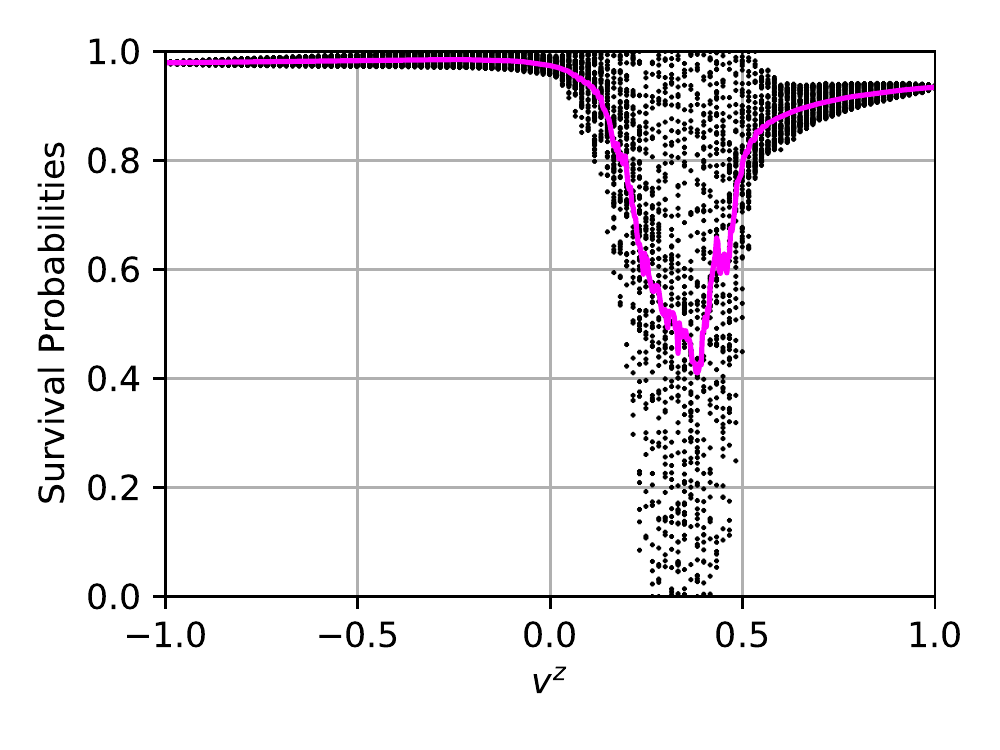}
\llap{\parbox[b]{1.3in}{\small (a)\\\rule{0ex}{0.95in}}} }
\subfigure{\includegraphics[width=0.22\textwidth]{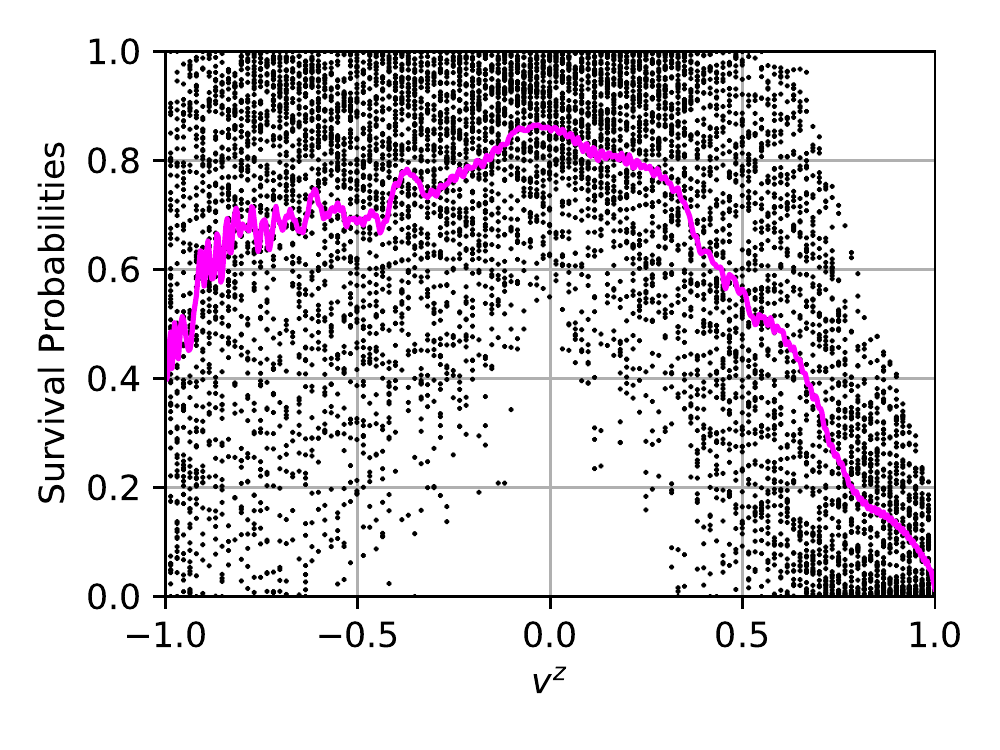}
\llap{\parbox[b]{0.3in}{\small (f)\\\rule{0ex}{0.95in}}} }\\
\subfigure{\includegraphics[width=0.22\textwidth]{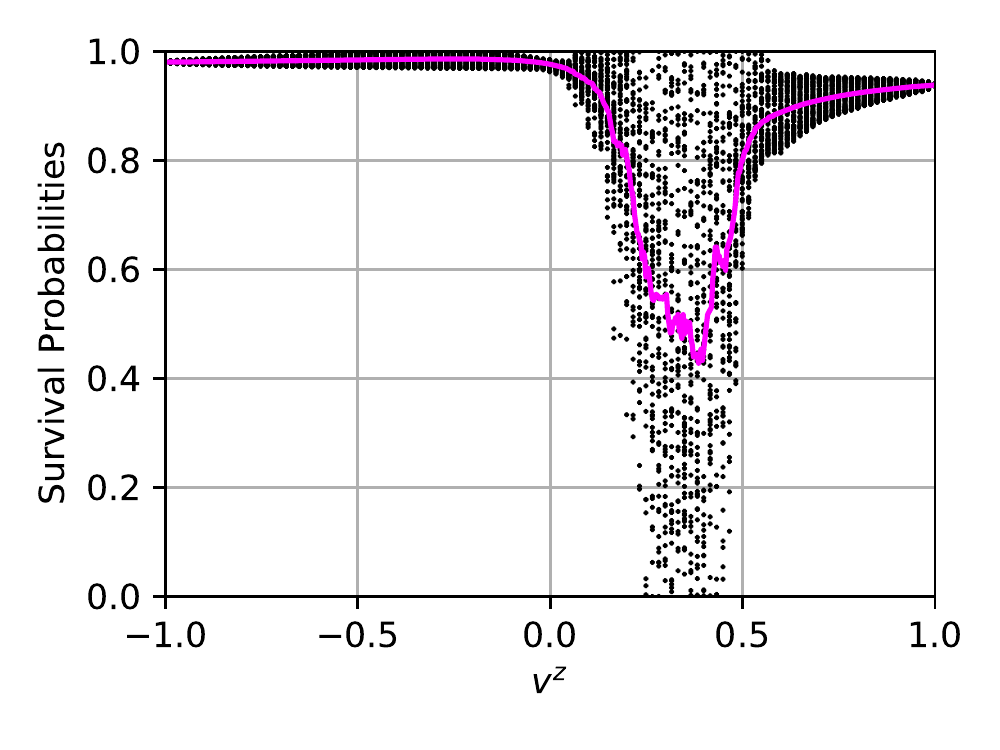}
\llap{\parbox[b]{1.3in}{\small (b)\\\rule{0ex}{0.95in}}} }
\subfigure{\includegraphics[width=0.22\textwidth]{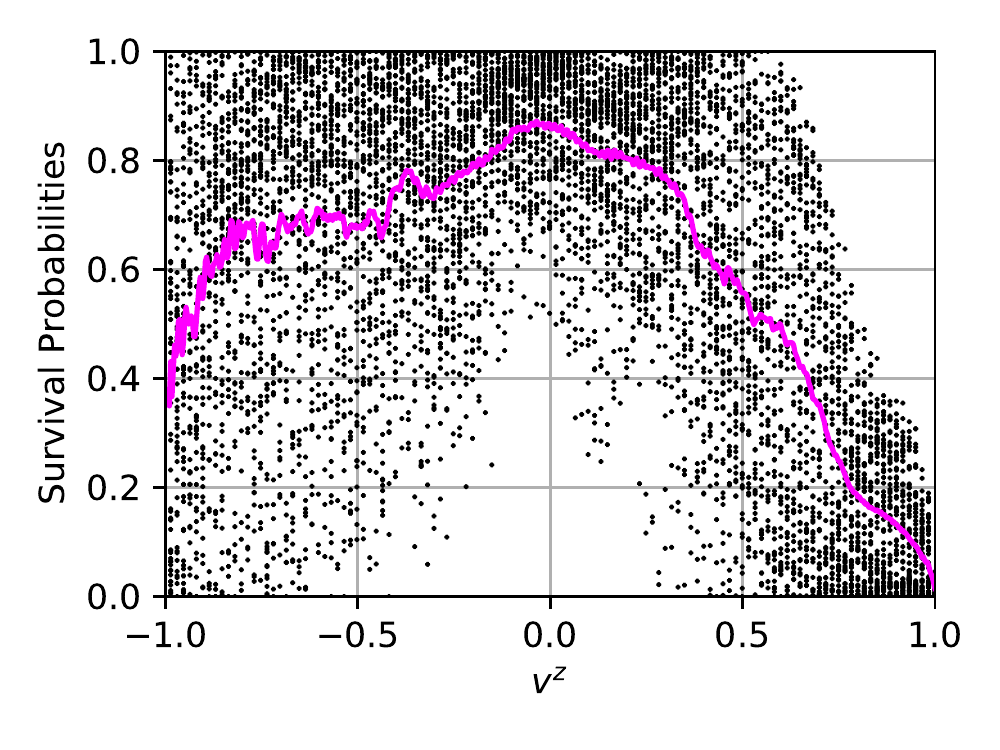}
\llap{\parbox[b]{0.3in}{\small (g)\\\rule{0ex}{0.95in}}} }\\
\subfigure{\includegraphics[width=0.22\textwidth]{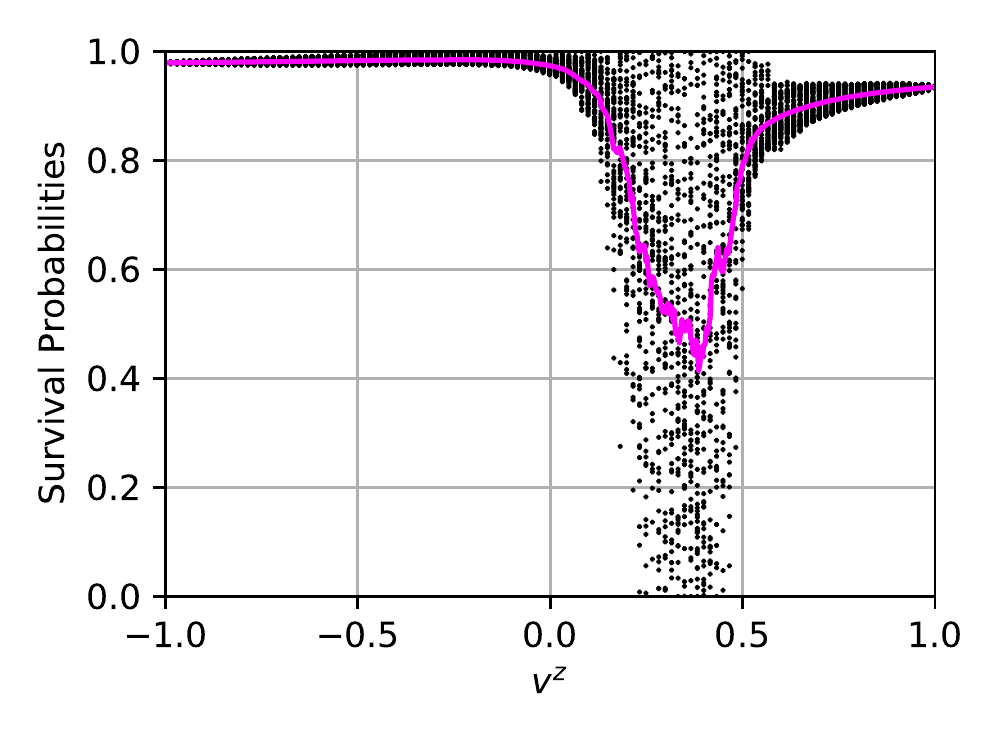}
\llap{\parbox[b]{1.3in}{\small (c)\\\rule{0ex}{0.95in}}} }
\subfigure{\includegraphics[width=0.22\textwidth]{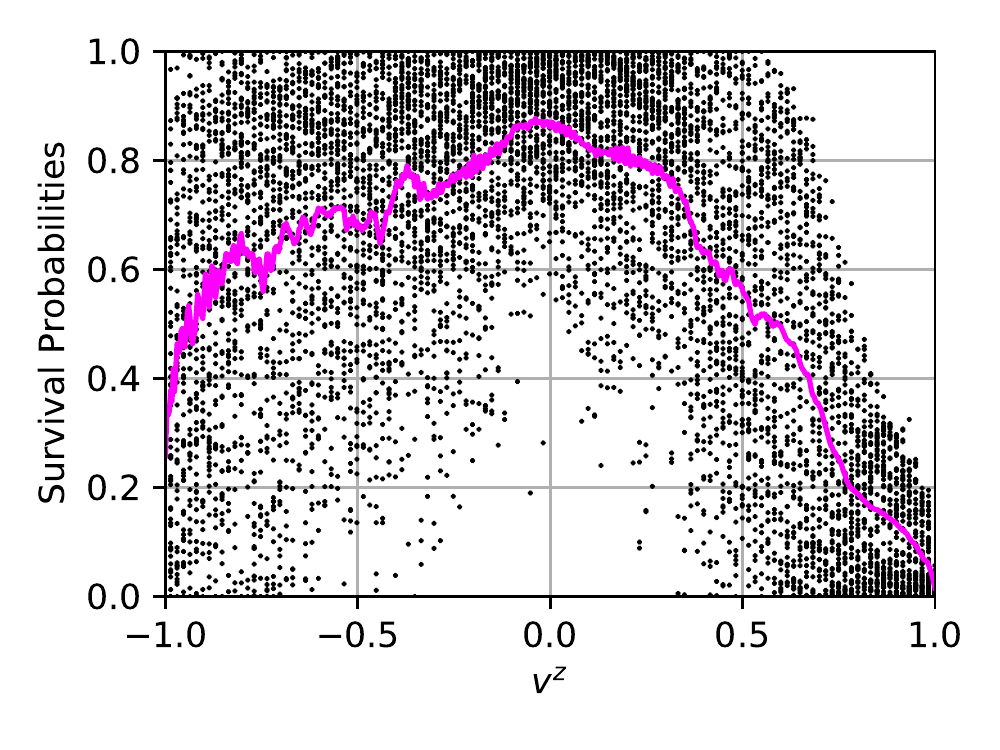}
\llap{\parbox[b]{0.3in}{\small (h)\\\rule{0ex}{0.95in}}} }\\
\subfigure{\includegraphics[width=0.22\textwidth]{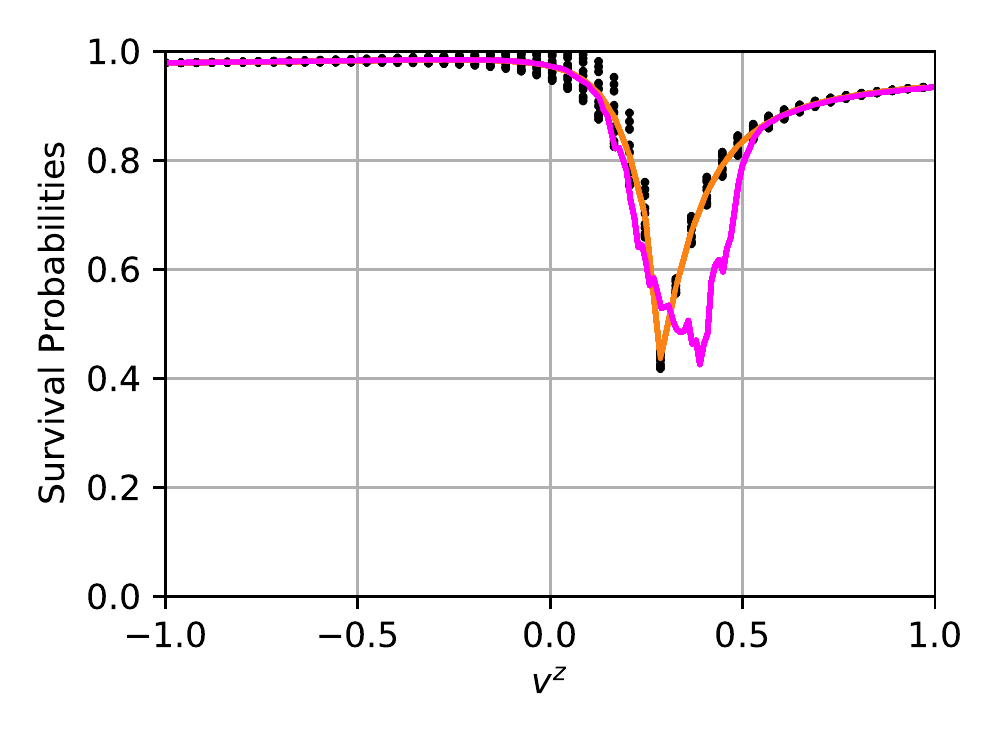}
\llap{\parbox[b]{1.3in}{\small (d)\\\rule{0ex}{0.95in}}} }
\subfigure{\includegraphics[width=0.22\textwidth]{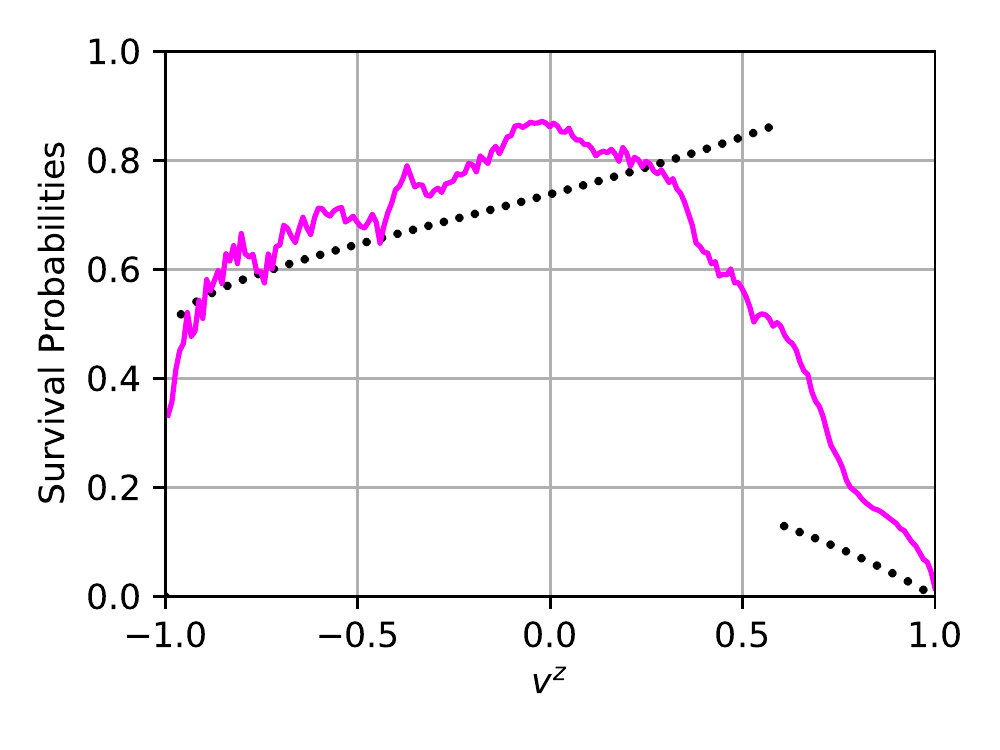}
\llap{\parbox[b]{0.3in}{\small (i)\\\rule{0ex}{0.95in}}} }\\
\subfigure{\includegraphics[width=0.22\textwidth]{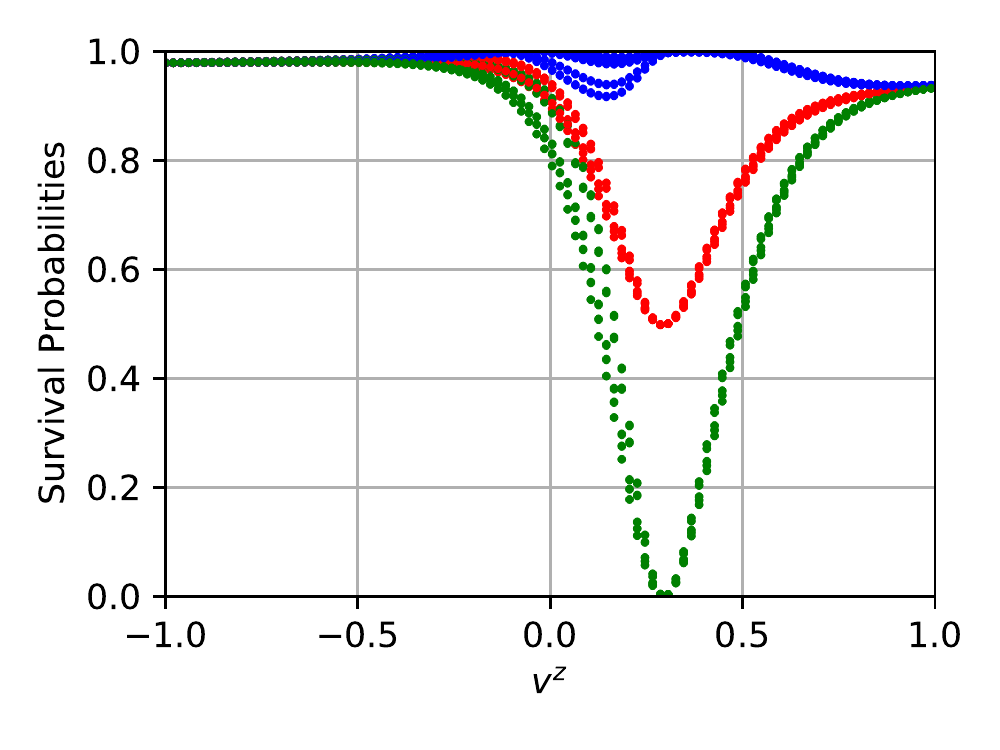}
\llap{\parbox[b]{1.3in}{\small (e)\\\rule{0ex}{0.95in}}} }
\subfigure{\includegraphics[width=0.22\textwidth]{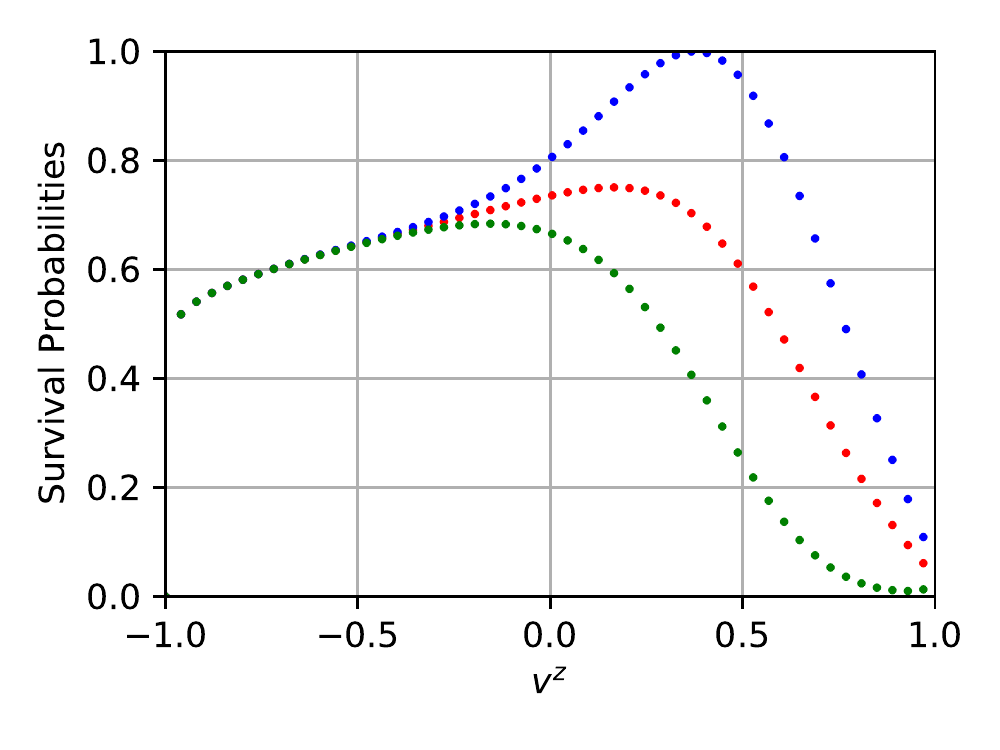}
\llap{\parbox[b]{0.3in}{\small (j)\\\rule{0ex}{0.95in}}} }
	\caption{\label{fig:other_spectra} 
	Comparison of the survival probabilities from the stationary
	solutions with the numerical results for the $\nu$ELN distributions of cases A and B. 
	Symbols at the same $v^z$ are for different azimuthal angles.
	In panels (a), (b), and (c), snapshots of the numerical results are shown 
	for $t=1.6$, 1.8, and 2~km, respectively, for case A. The magenta curves are the survival probabilities 
	averaged over the azimuthal angle. In panel (d), the symbols show the stationary solution for case A, 
	the orange curve is obtained by averaging these symbols over the azimuthal angle,
	and the magenta curve is the same as that in panel (c). In panel (e), the symbols show
	the survival probabilities (red: mean, blue and green: limits) including the approximate effect 
	of nonadiabatic evolution for the stationary solution for case A. Panels (f)--(j) are similar to
	panels (a)--(e), but for case B.}
\end{figure}

\begin{figure}[ht!]
\centering
\includegraphics[width=0.48\textwidth]{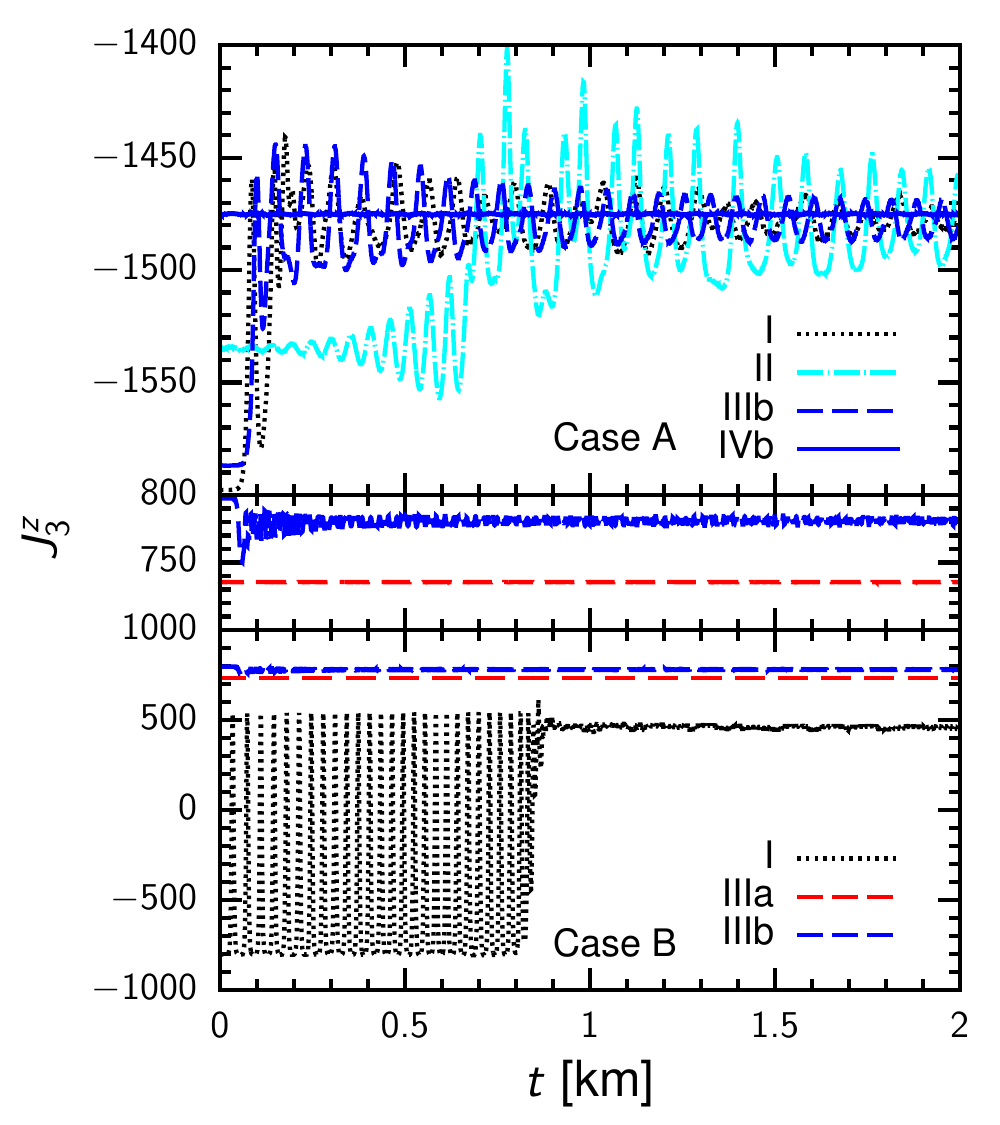}
	\caption{\label{fig:convergence3} Evolution of $J^z_3$ for the candidate
	stationary solutions for the $\nu$ELN distributions of cases A and B
	after they are perturbed. For case B, the difference between types IIIa and IIIb is highlighted in the middle panel.}
\end{figure}

Figure~\ref{fig:other_spectra} shows that an approximately stationary state has been reached by
$t=2$~km for the explicit evolution calculations for both cases A and B, and compares the 
survival probabilities from the stationary solutions with the numerical results. 
For case A, we find three nontrivial candidate stationary solutions of types II, IIIb, and IVb. 
Only type IVb is stable (see Fig.~\ref{fig:convergence3}), and the corresponding average survival
probabilities match the numerical results very well away from the zero-crossing at $v^z\approx 0.38$
[see Fig.~\ref{fig:other_spectra}(d)].
The trend and range of the numerical results can also be explained by including the
approximate effect of nonadiabatic evolution based on this solution with
\begin{align}
\epsilon_{\rm eff}(\mathbf{v})=\text{erf}\left[2H_{3,t_0}'(\mathbf{v})/\mu\right]
\end{align}
replacing Eq.~(\ref{eq:cf_epsilon2}) [see Fig.~\ref{fig:other_spectra}(e)].
For case B, we find two nontrivial candidate stationary solutions of types IIIa and IIIb. 
Only type IIIa is stable (see Fig.~\ref{fig:convergence3}). The corresponding average survival 
probabilities qualitatively follow the numerical results away from the zero-crossing at 
$v^z\approx 0.6$ [see Fig.~\ref{fig:other_spectra}(i)].
The trend and range of the numerical results for $v^z\gtrsim 0.5$ can also be explained by 
including the approximate effect of nonadiabatic evolution based on this solution with
\begin{align}
\epsilon_{\rm eff}(\mathbf{v})=\text{erf}\left[3H_{3,t_0}'(\mathbf{v})/\mu\right]
\end{align}
replacing Eq.~(\ref{eq:cf_epsilon2}) [see Fig.~\ref{fig:other_spectra}(j)].

The flavor evolution for the $\nu$ELN distribution of case B differs from
all the other cases considered in that it is associated with a sign flip of $J^z_3$.
As $J^z_3$ changes from the initial value of $-803\,\mathrm{km}^{-1}$ to
$\approx 450\,\mathrm{km}^{-1}$ for the approximately stationary state
(see Fig.~\ref{fig:convergence3}),
neutrinos with a wide range of $v^z$ undergo nonadiabatic evolution. This effect
cannot be captured by our approximate treatment of nonadiabatic evolution, which is 
appropriate only for a limited range of $v^z$ around the zero-crossing.
We note, however, that for the likely inhomogeneous astrophysical environments,
the dominant instability associated with the $\nu$ELN distribution of case B 
grows differently on different spatial scales \cite{yi2019dispersion,martin2020dynamic}. Therefore,
a realistic treatment of this case must address both the temporal and spatial 
flavor evolution of neutrinos, which is beyond our scope here.

\bibliographystyle{elsarticle-num}
\bibliography{reference}

\end{document}